\documentclass[journal]{vgtc}         
\usepackage{dblfloatfix}

\onlineid{6581}

\vgtccategory{Research}

\vgtcpapertype{theory/model}

\title{A Survey of Designs for Combined 2D\raisebox{1.5pt}{+}3D Visual Representations}

\author{%
  \authororcid{Jiayi Hong}{0000-0002-1332-5045},
  \authororcid{Rostyslav Hnatyshyn}{0009-0006-0510-1152},
  Ebrar A. D. Santos,
  \authororcid{Ross Maciejewski}{0000-0001-8803-6355}, and 
  \authororcid{Tobias Isenberg}{0000-0001-7953-8644}
}

\authorfooter{
  \item
  	J.\ Hong (\raisebox{-.5pt}{\includegraphics[height=6pt]{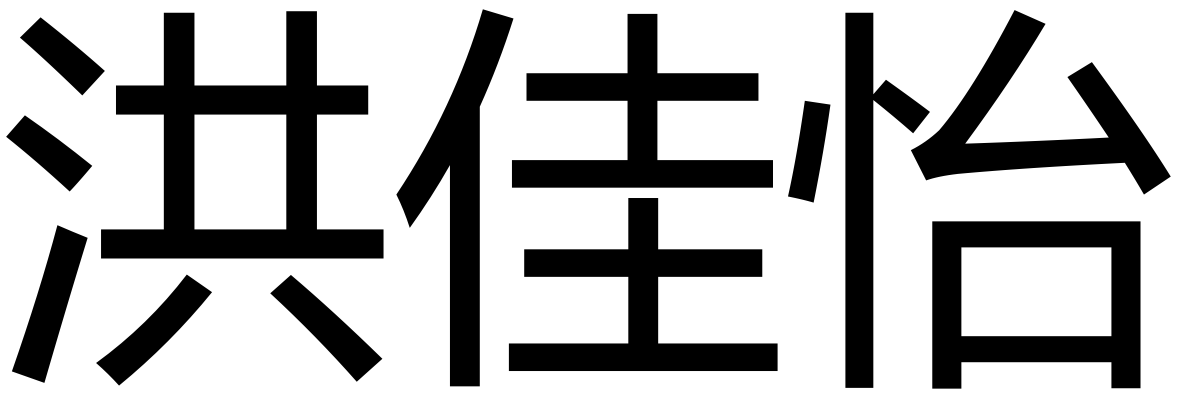}}), R.\ Hnatyshyn, and R.\ Maciejewski are with Arizona State University, USA.
        E-mail: {\{jhong76\,$|$\,rhnatysh\,$|$\,rmacieje\}}@asu.edu.
  \item
  	T.\ Isenberg is with Universit\'e Paris-Saclay, CNRS, Inria, France. 
        E-mail: tobias.isenberg@inria.fr.
}

\abstract{%
We examine visual representations of data that make use of combinations of both 2D and 3D data mappings. Combining 2D and 3D representations is a common technique that allows viewers to understand multiple facets of the data with which they are interacting.
While 3D representations focus on the spatial character of the data or the dedicated 3D data mapping, 2D representations often show abstract data properties and take advantage of the unique benefits of mapping to a plane.
Many systems have used unique combinations of both types of data mappings effectively. Yet there are no systematic reviews of the methods in linking 2D and 3D representations.
We systematically survey the relationships between 2D and 3D visual representations in major visualization publications---IEEE VIS, IEEE TVCG, and EuroVis---from 2012 to 2022. We closely examined 105 papers where 2D and 3D representations are connected visually, interactively, or through animation. 
These approaches are designed based on their visual environment, the relationships between their visual representations, and their possible layouts. Through our analysis, we introduce a design space as well as provide design guidelines for effectively linking 2D and 3D visual representations.
}


\teaser{
  \centering
  \includegraphics[width=\linewidth]{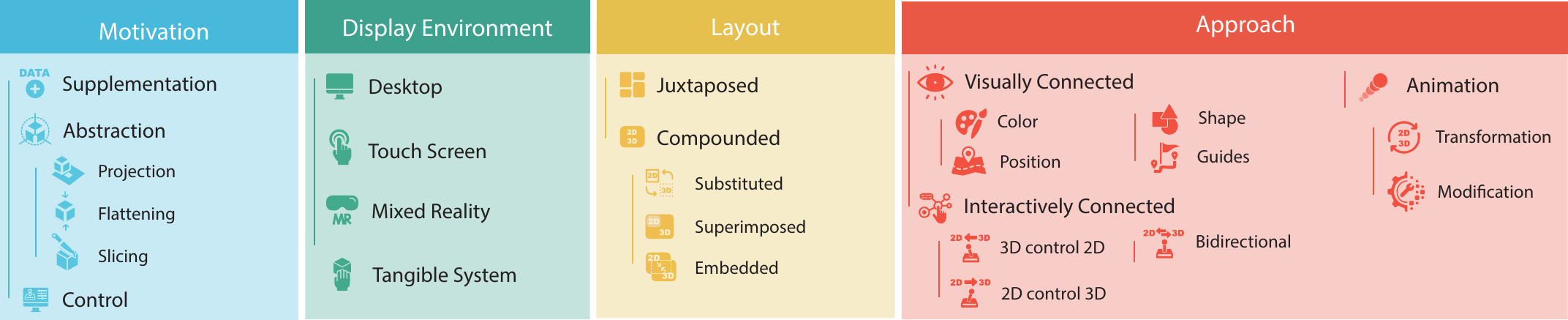}%\vspace{-1ex}
  \caption{Overview of our proposed design space for combined 2D+3D representations.}
  \label{fig:overview_design_space}
}

\graphicspath{{figs/}{figures/}{pictures/}{images/}{./}} % where to search for the images
\usepackage[svgnames]{xcolor} % needed again and not only at end of praeamble as by style due to upcoming definitions
\usepackage{tabu}                      % only used for the table example
\usepackage{booktabs}                  % only used for the table example
\usepackage{lipsum}                    % used to generate placeholder text
\usepackage{mwe}                       % used to generate placeholder figures
\usepackage[normalem]{ulem}
\usepackage{microtype}
\usepackage{graphicx}
\usepackage{textcomp}
\usepackage{mathptmx} 
\usepackage{enumitem}
\usepackage{ccicons}
\usepackage{multirow}
\usepackage{soul}
\usepackage{amsmath}
\usepackage{amssymb}
 
\usepackage{cuted}

\newcommand{\eg}{e.\,g.}
\newcommand{\ie}{i.\,e.}

\newcommand{\jh}[1]{\textcolor{black}{#1}}

\newcommand\redsout{\bgroup\markoverwith{\textcolor{red}{\rule[0.5ex]{2pt}{0.4pt}}}\ULon}

 \renewcommand{\redsout}[1]{}

\definecolor{display-bg}{HTML}{c0e1dd}
\definecolor{approach-bg}{HTML}{f8d3ca}
\definecolor{layout-bg}{HTML}{f8edd1}
\definecolor{motivation-bg}{HTML}{bbe5f1}

\newcommand{\displaybg}[1]{\begingroup{}\setul{-1.8ex}{2ex}\setulcolor{display-bg}\ul{#1}\endgroup}
\newcommand{\approachbg}[1]{\begingroup{}\setul{-1.8ex}{2ex}\setulcolor{approach-bg}\ul{#1}\endgroup}
\newcommand{\layoutbg}[1]{\begingroup{}\setul{-1.8ex}{2ex}\setulcolor{layout-bg}\ul{#1}\endgroup}
\newcommand{\motivationbg}[1]{\begingroup{}\setul{-1.8ex}{2ex}\setulcolor{motivation-bg}\ul{#1}\endgroup}

\newlength{\iconheight}
\newlength{\iconverticaloffset}
\newlength{\iconspace}
\DeclareRobustCommand{\desktopicon}{%
  \begingroup\normalfont
\raisebox{\iconverticaloffset}[0pt][0pt]{\includegraphics[height=\iconheight]{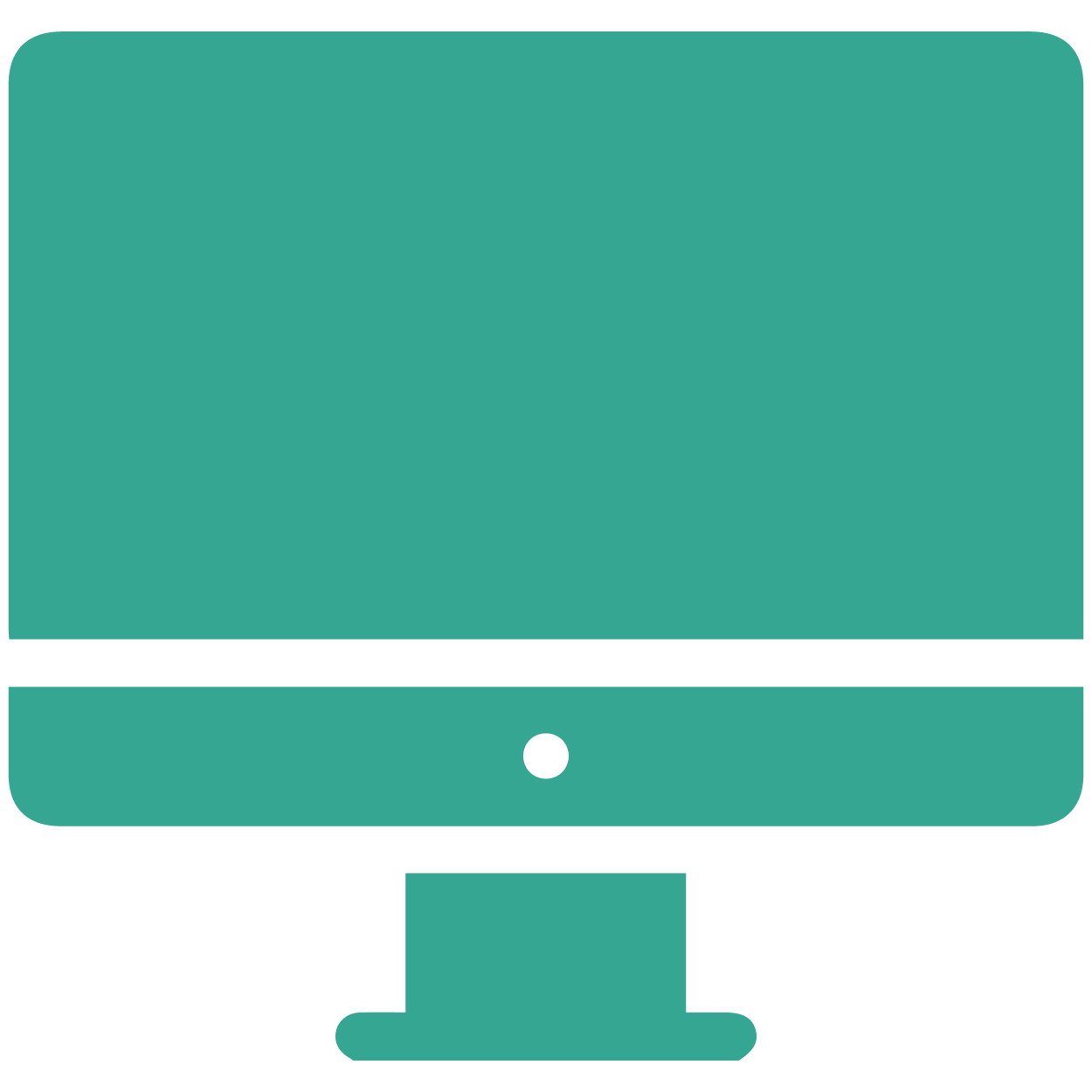}}%
  \endgroup
}
\DeclareRobustCommand{\desktop}{\desktopicon\hspace{\iconspace}~\displaybg{Desktop}}

\DeclareRobustCommand{\mricon}{%
  \begingroup\normalfont
\raisebox{\iconverticaloffset}[0pt][0pt]%
{\includegraphics[height=\iconheight]{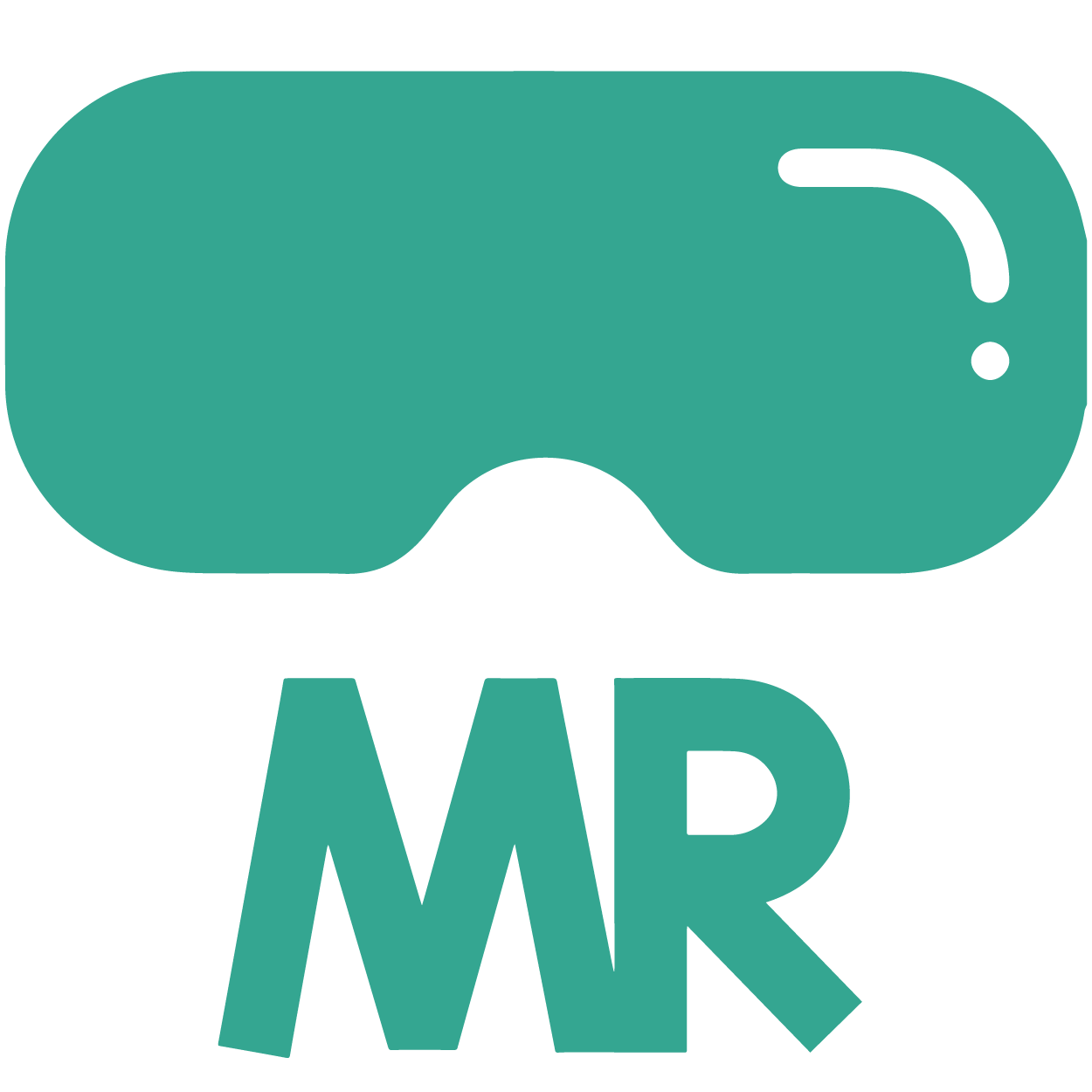}}%
  \endgroup
}
\DeclareRobustCommand{\mr}{\mricon\hspace{\iconspace}~\displaybg{Mixed Reality}}

\DeclareRobustCommand{\touchicon}{%
  \begingroup\normalfont
\raisebox{\iconverticaloffset}[0pt][0pt]%
{\includegraphics[height=\iconheight]{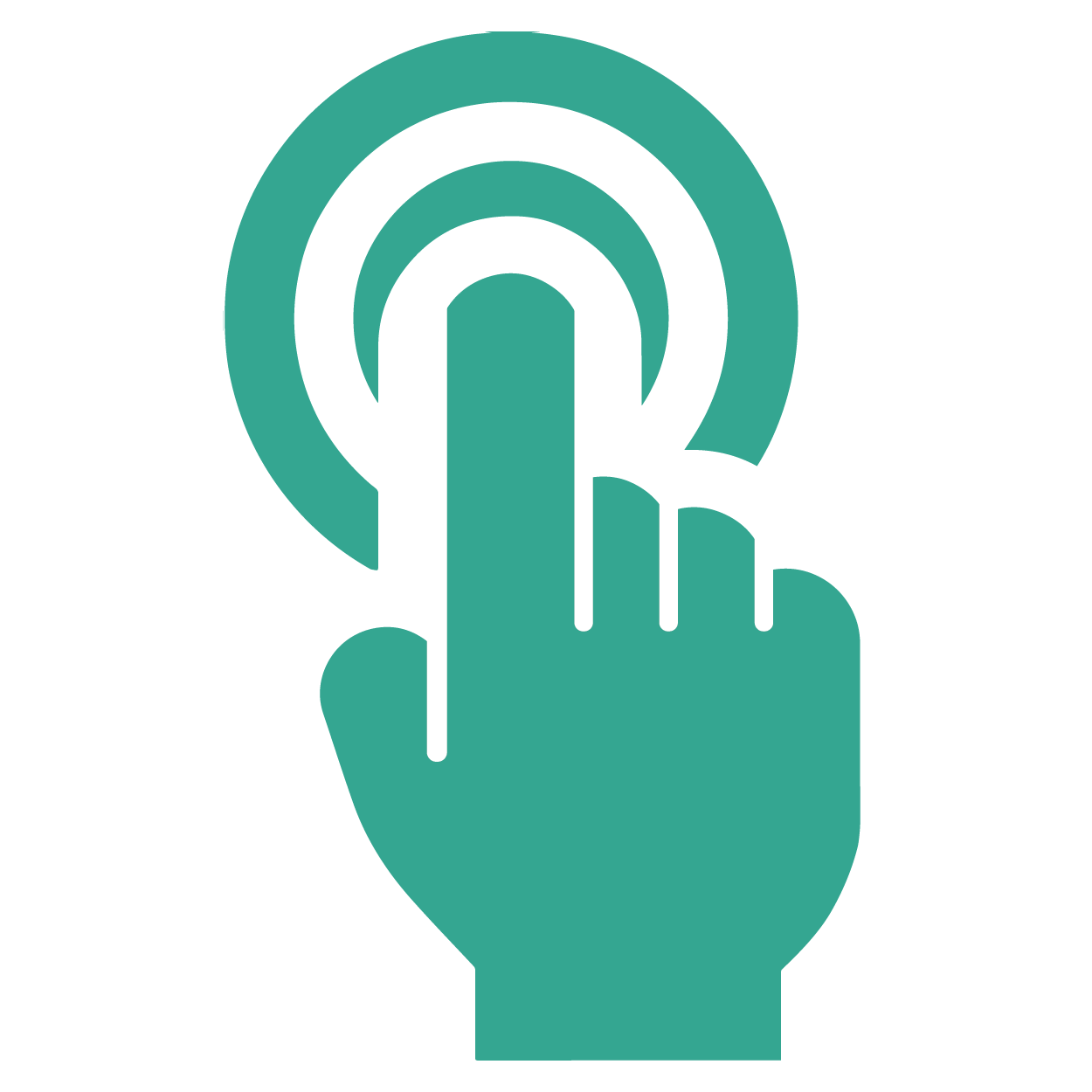}}%
  \endgroup
}
\DeclareRobustCommand{\touch}{\touchicon\hspace{\iconspace}~\displaybg{Touch Screen}}

\DeclareRobustCommand{\tangibleicon}{%
  \begingroup\normalfont
\raisebox{\iconverticaloffset}[0pt][0pt]%
{\includegraphics[height=\iconheight]{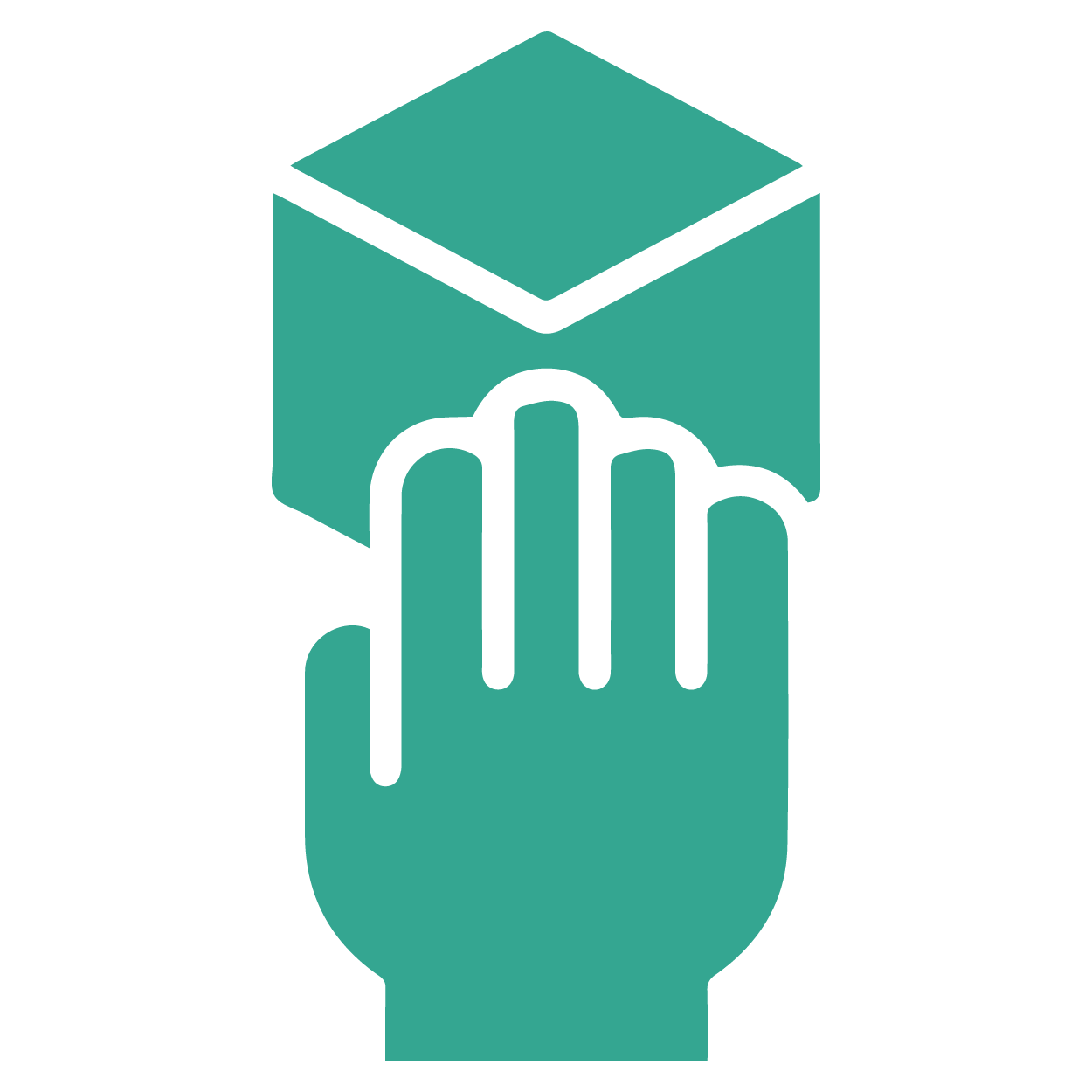}}%
  \endgroup
}
\DeclareRobustCommand{\tangible}{\tangibleicon\hspace{\iconspace}~\displaybg{Tangible System}}
\DeclareRobustCommand{\tangibleP}{\tangibleicon\hspace{\iconspace}~\displaybg{Tangible Systems}}

\DeclareRobustCommand{\visualicon}{%
  \begingroup\normalfont
\raisebox{\iconverticaloffset}[0pt][0pt]%
{\includegraphics[height=\iconheight]{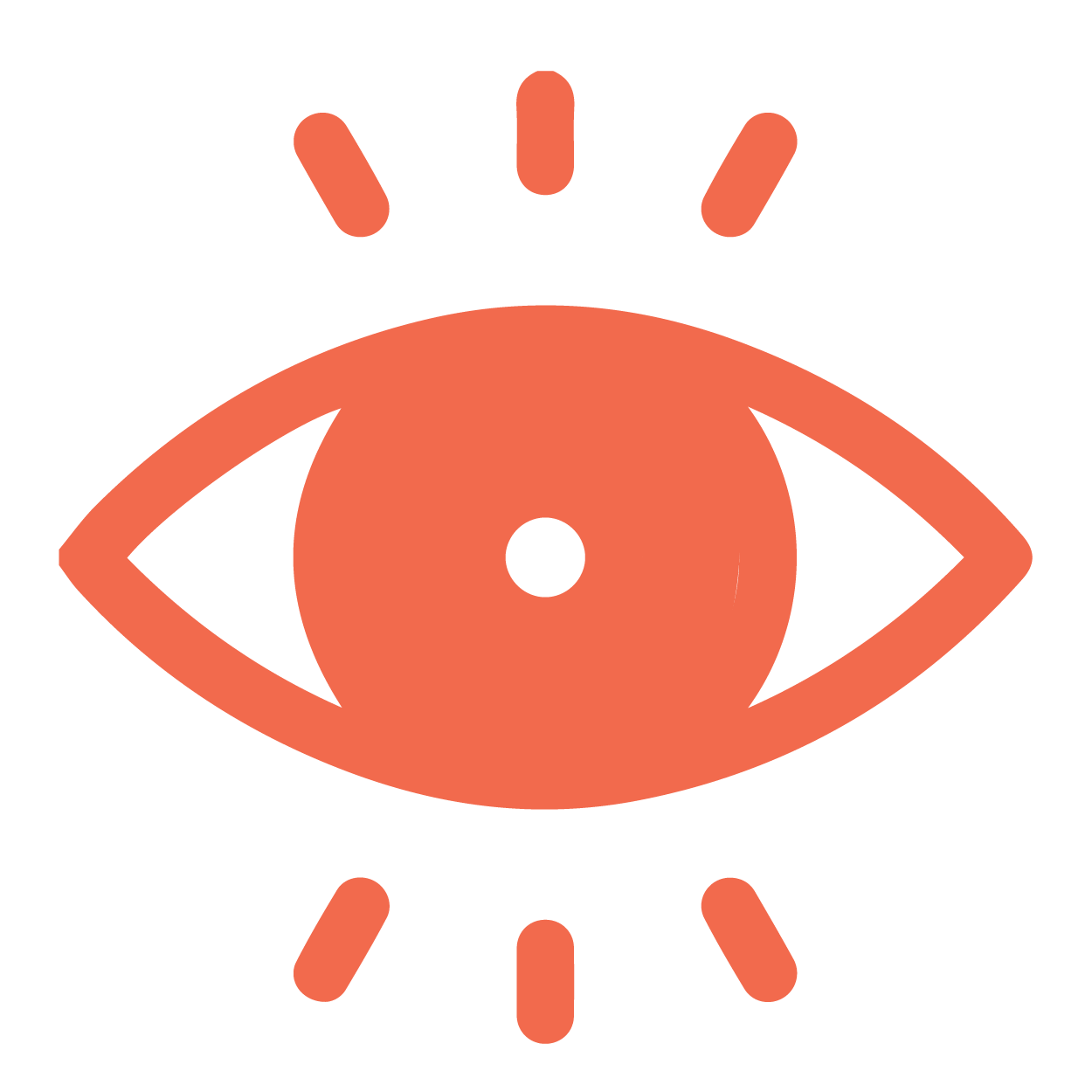}}%
  \endgroup
}
\DeclareRobustCommand{\visual}{\visualicon\hspace{\iconspace}~\approachbg{Visually Connected}}

\DeclareRobustCommand{\coloricon}{%
  \begingroup\normalfont
\raisebox{\iconverticaloffset}[0pt][0pt]%
{\includegraphics[height=\iconheight]{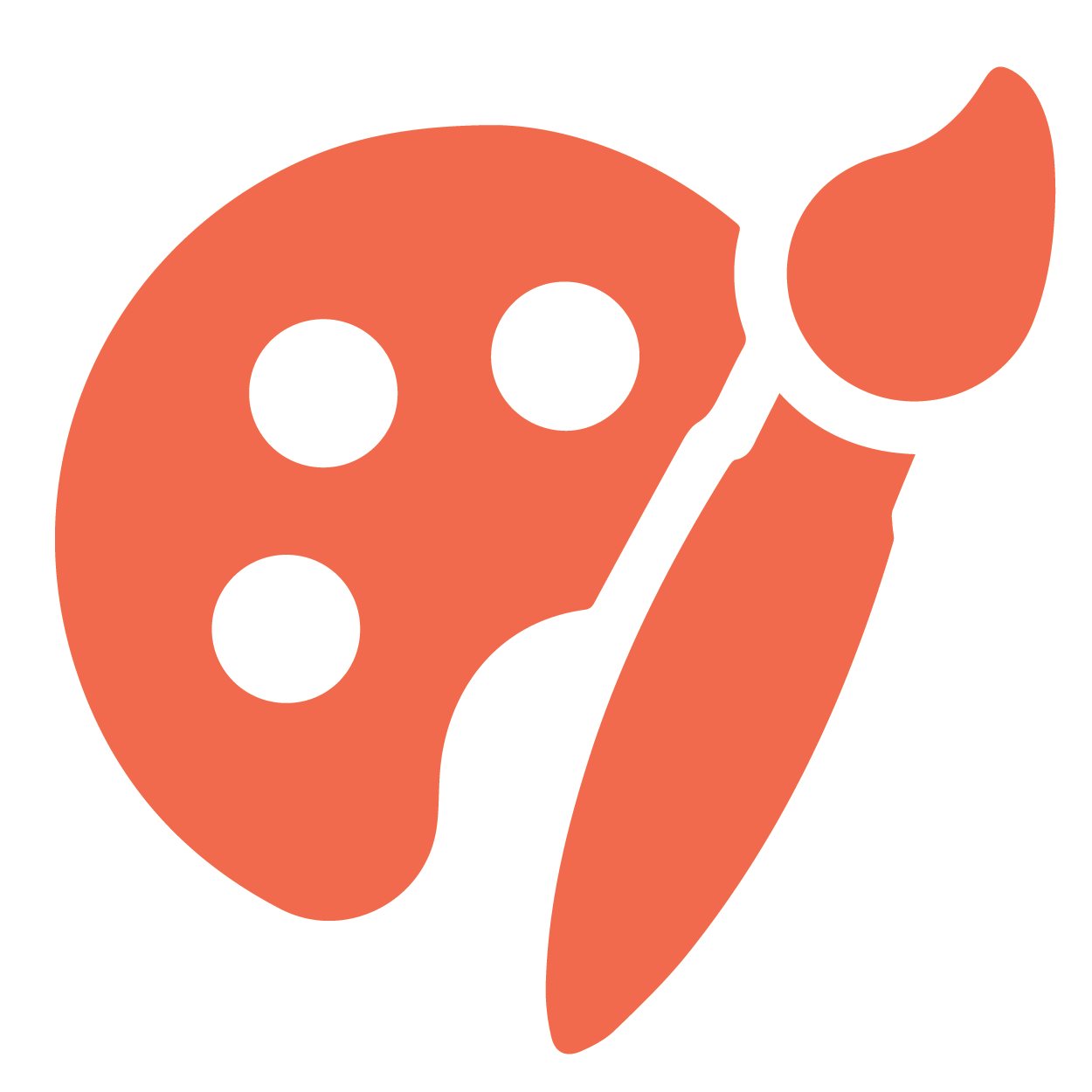}}%
  \endgroup
}
\DeclareRobustCommand{\colorcn}{\visualicon~\coloricon\hspace{\iconspace}~\approachbg{Color}}

\DeclareRobustCommand{\positionicon}{%
  \begingroup\normalfont
\raisebox{\iconverticaloffset}[0pt][0pt]%
{\includegraphics[height=\iconheight]{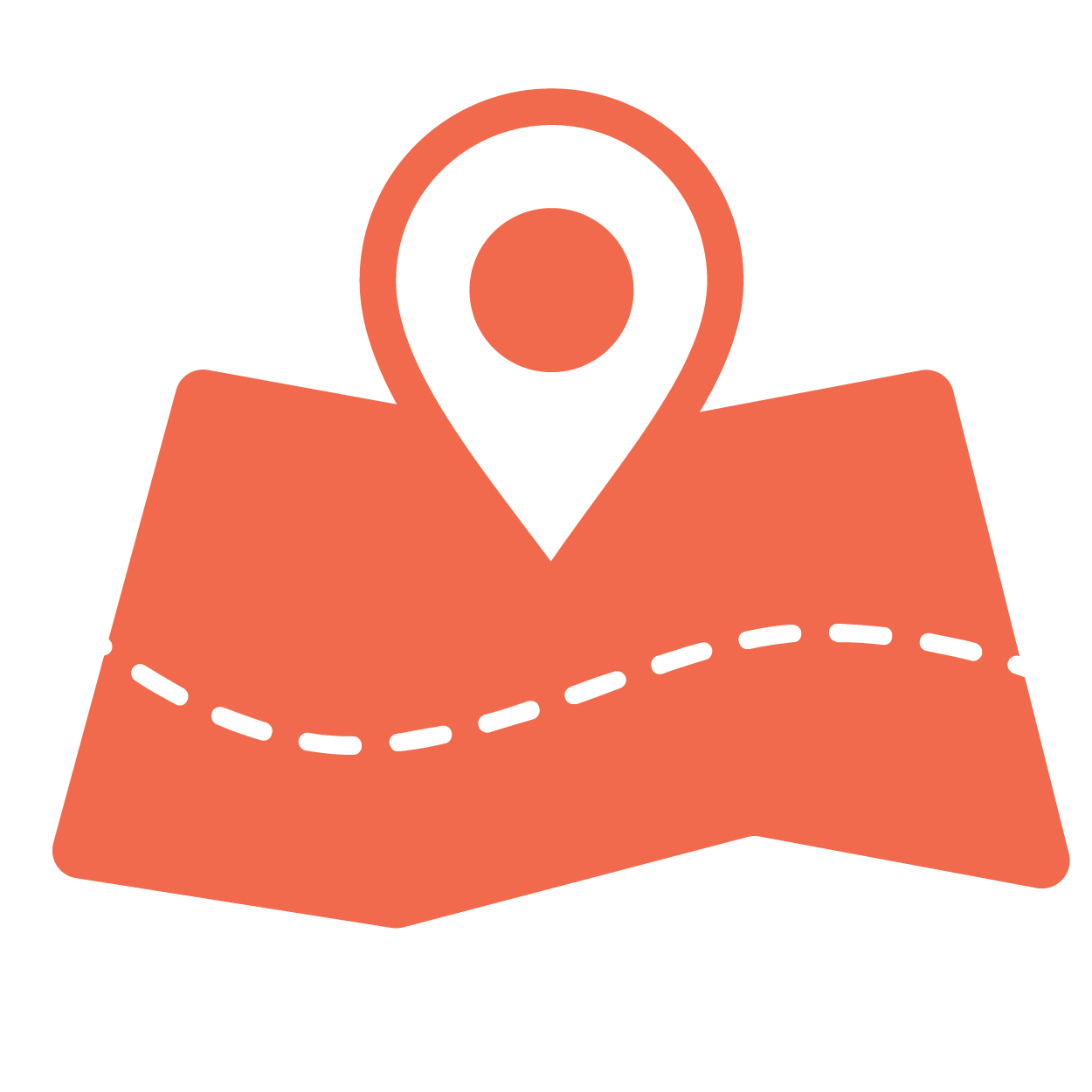}}%
  \endgroup
}
\DeclareRobustCommand{\positioncn}{\visualicon~\positionicon\hspace{\iconspace}~\approachbg{Position}}

\DeclareRobustCommand{\shapeicon}{%
  \begingroup\normalfont
\raisebox{\iconverticaloffset}[0pt][0pt]%
{\includegraphics[height=\iconheight]{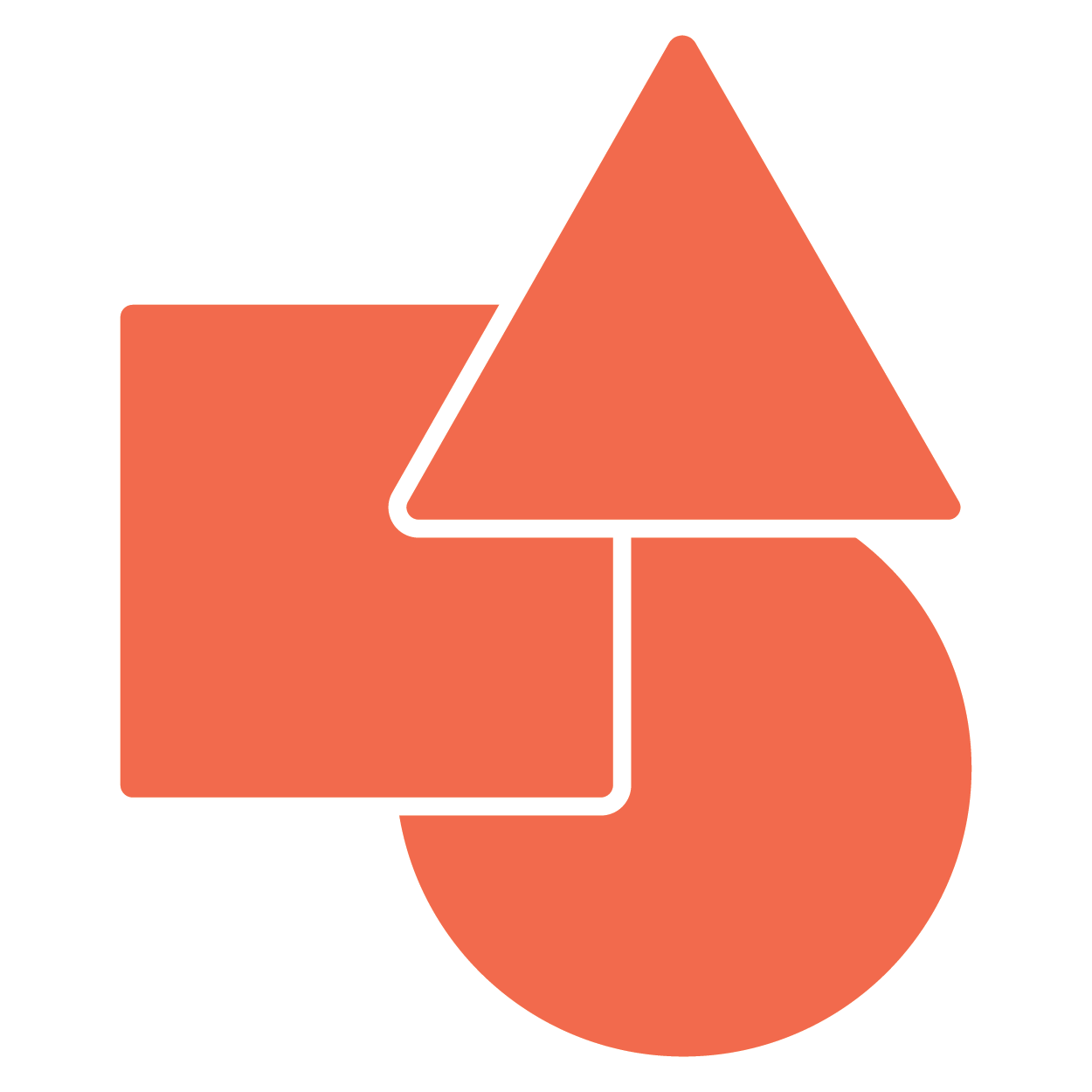}}%
  \endgroup
}
\DeclareRobustCommand{\shapecn}{\visualicon~\shapeicon\hspace{\iconspace}~\approachbg{Shape}}

\DeclareRobustCommand{\guideicon}{%
  \begingroup\normalfont
\raisebox{\iconverticaloffset}[0pt][0pt]%
{\includegraphics[height=\iconheight]{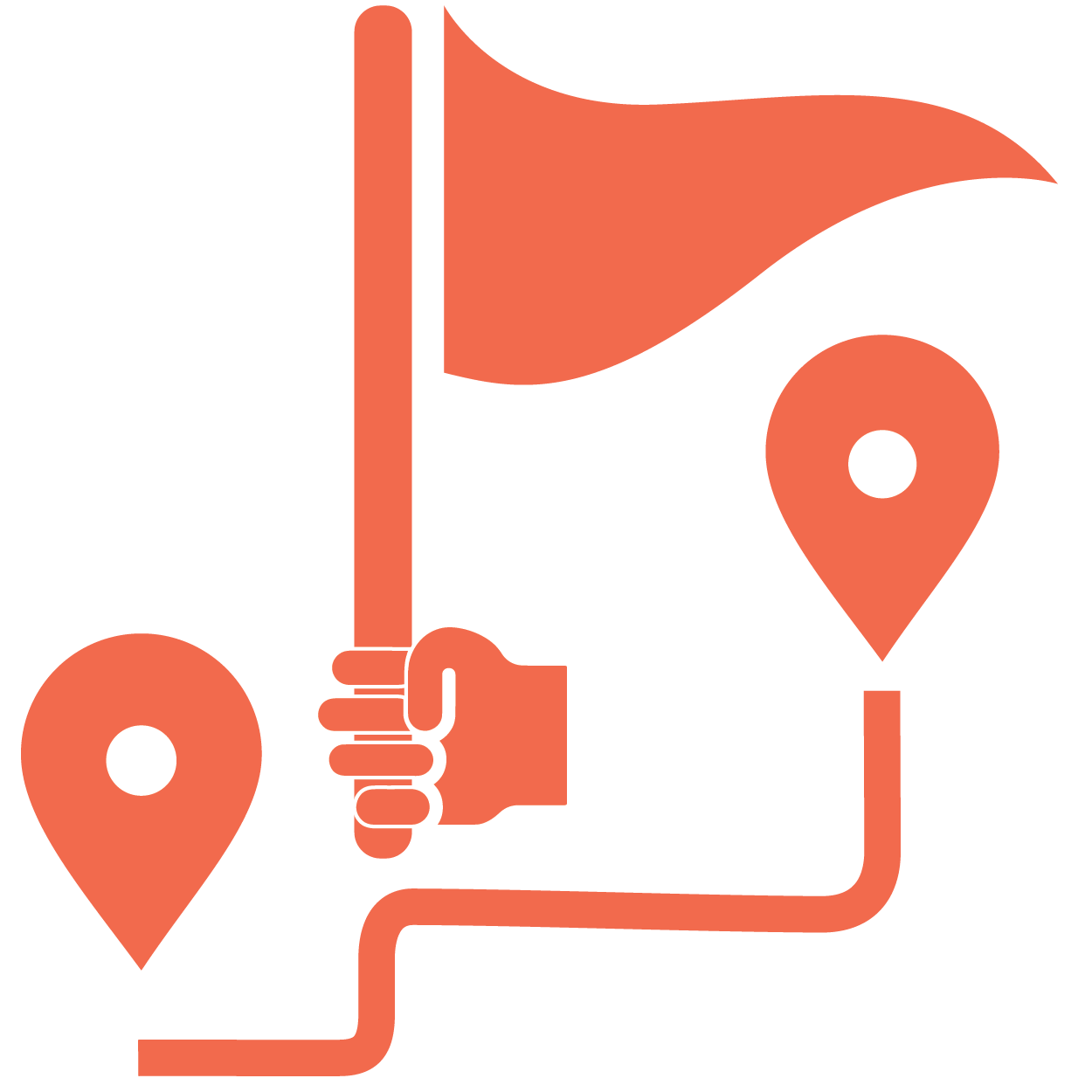}}%
  \endgroup
}
\DeclareRobustCommand{\guidecn}{\visualicon~\guideicon\hspace{\iconspace}~\approachbg{Guides}}

\DeclareRobustCommand{\interacticon}{%
  \begingroup\normalfont
\raisebox{\iconverticaloffset}[0pt][0pt]%
{\includegraphics[height=\iconheight]{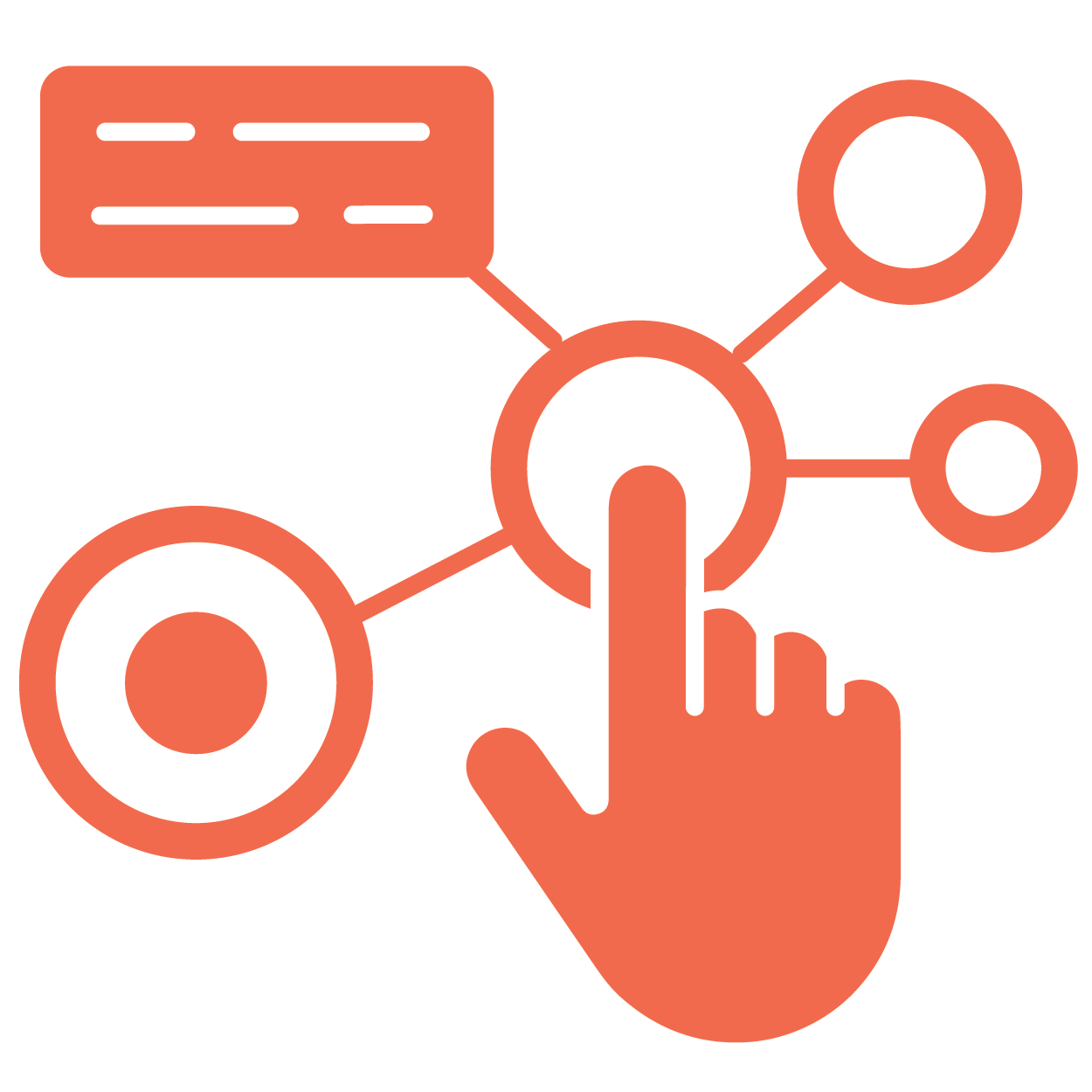}}%
  \endgroup
}
\DeclareRobustCommand{\interactive}{\interacticon\hspace{\iconspace}~\approachbg{Interactively Connected}}

\DeclareRobustCommand{\twocthreeicon}{%
  \begingroup\normalfont
\raisebox{\iconverticaloffset}[0pt][0pt]%
{\includegraphics[height=\iconheight]{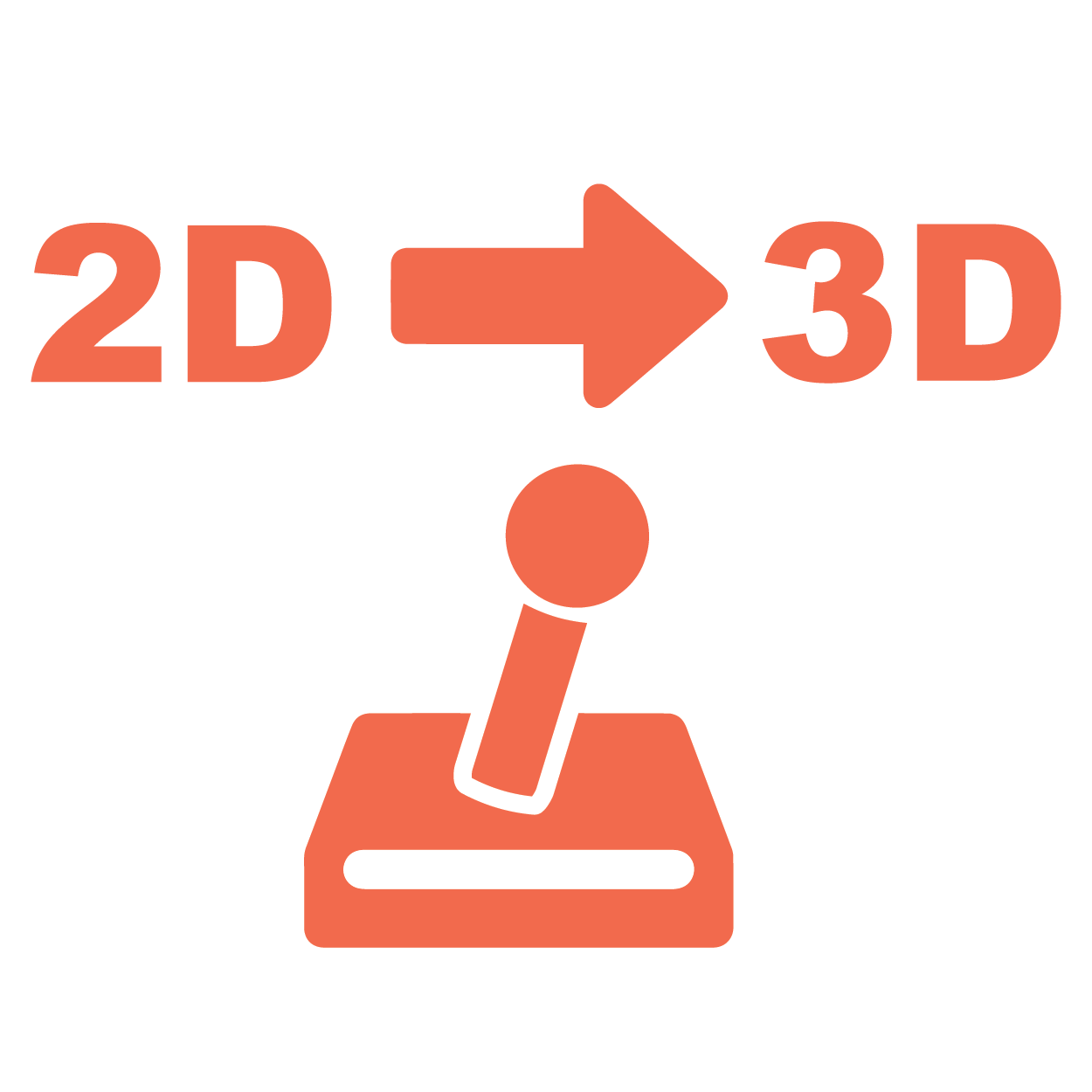}}%
  \endgroup
}
\DeclareRobustCommand{\twocthree}{\interacticon~\twocthreeicon\hspace{\iconspace}~\approachbg{2D Control 3D}}

\DeclareRobustCommand{\threectwoicon}{%
  \begingroup\normalfont
\raisebox{\iconverticaloffset}[0pt][0pt]%
{\includegraphics[height=\iconheight]{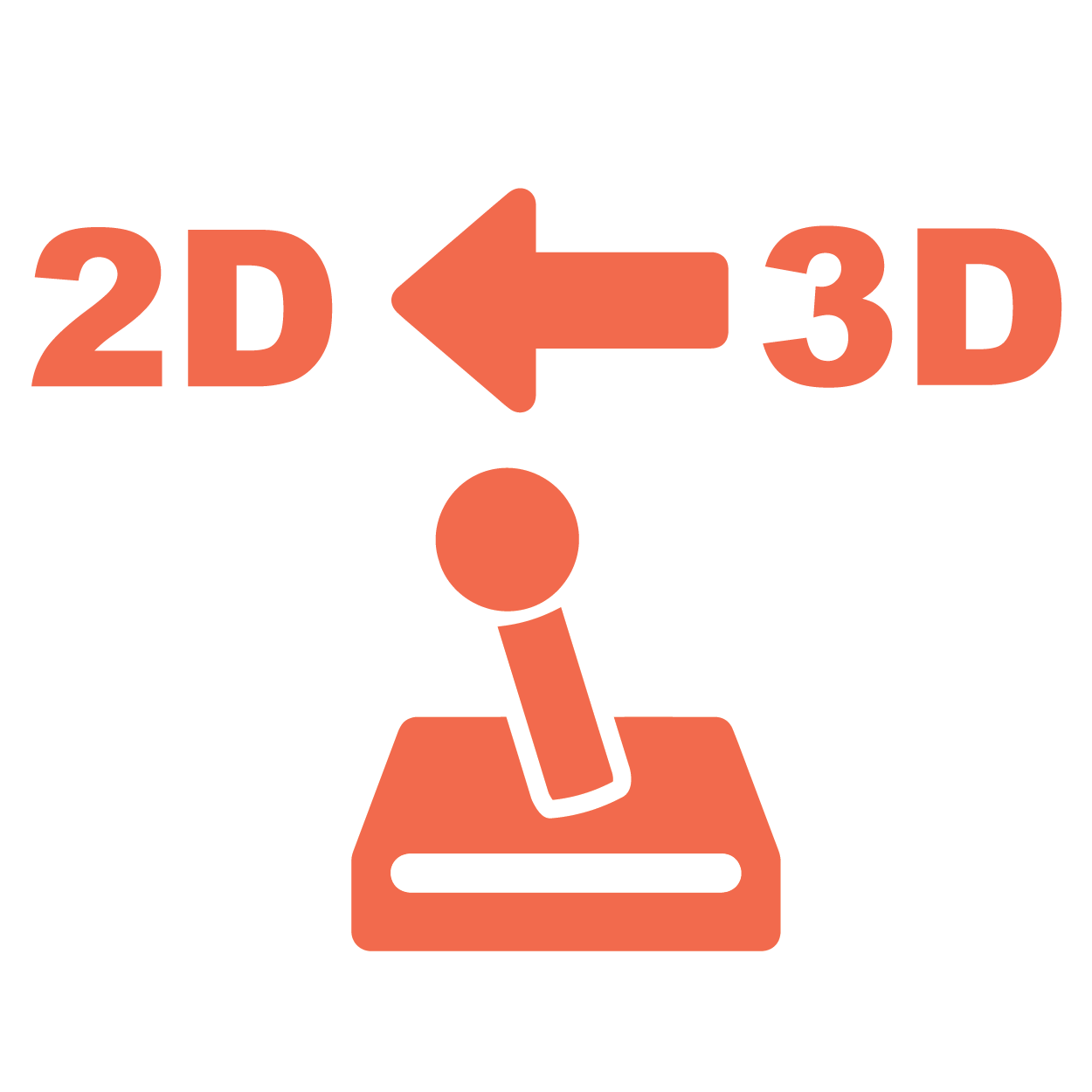}}%
  \endgroup
}
\DeclareRobustCommand{\threectwo}{\interacticon~\threectwoicon\hspace{\iconspace}~\approachbg{3D Control 2D}}

\DeclareRobustCommand{\bidirectionalicon}{%
  \begingroup\normalfont
\raisebox{\iconverticaloffset}[0pt][0pt]%
{\includegraphics[height=\iconheight]{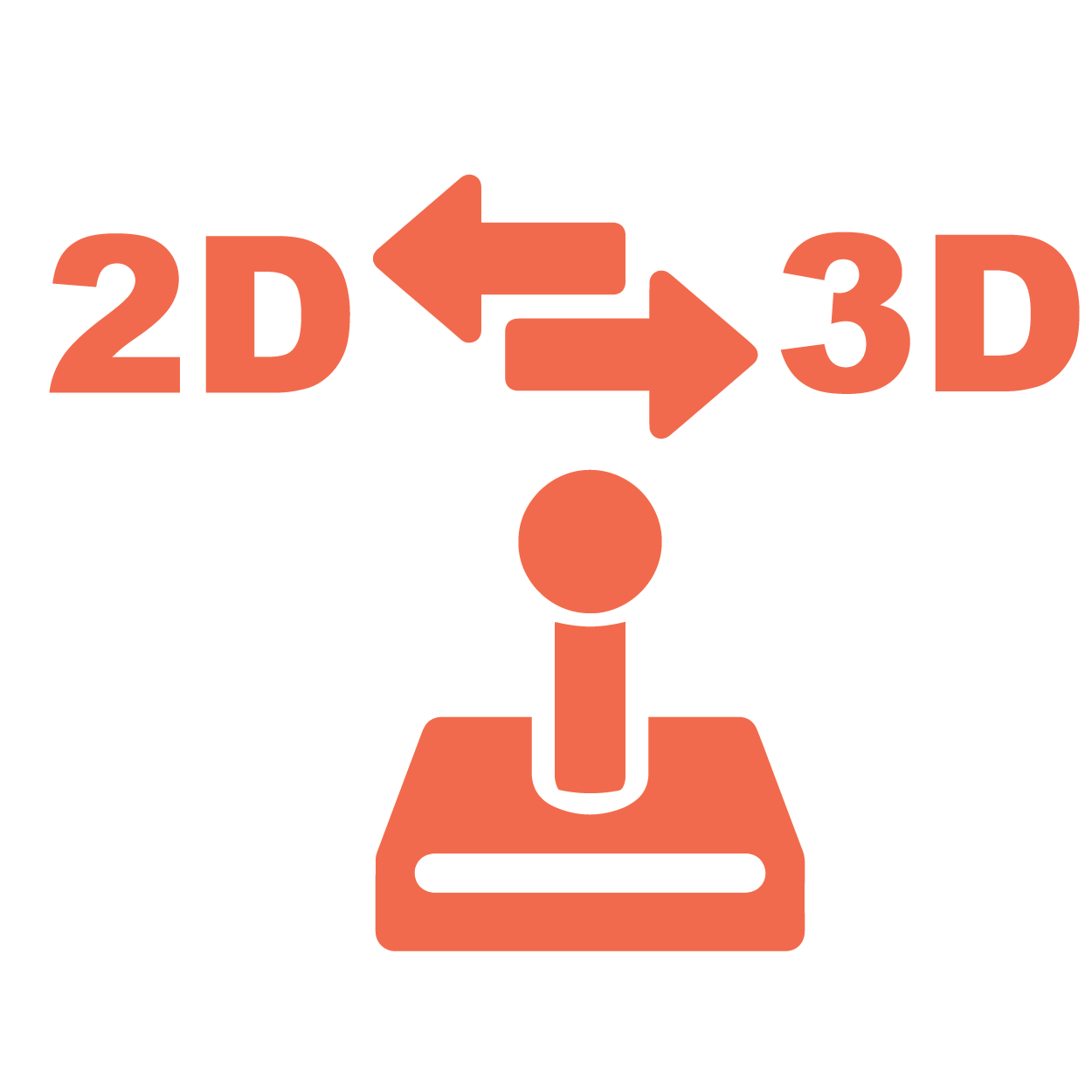}}%
  \endgroup
}
\DeclareRobustCommand{\bidirectional}{\interacticon~\bidirectionalicon\hspace{\iconspace}~\approachbg{Bidirectional}}

\DeclareRobustCommand{\animatedicon}{%
  \begingroup\normalfont
\raisebox{\iconverticaloffset}[0pt][0pt]%
{\includegraphics[height=\iconheight]{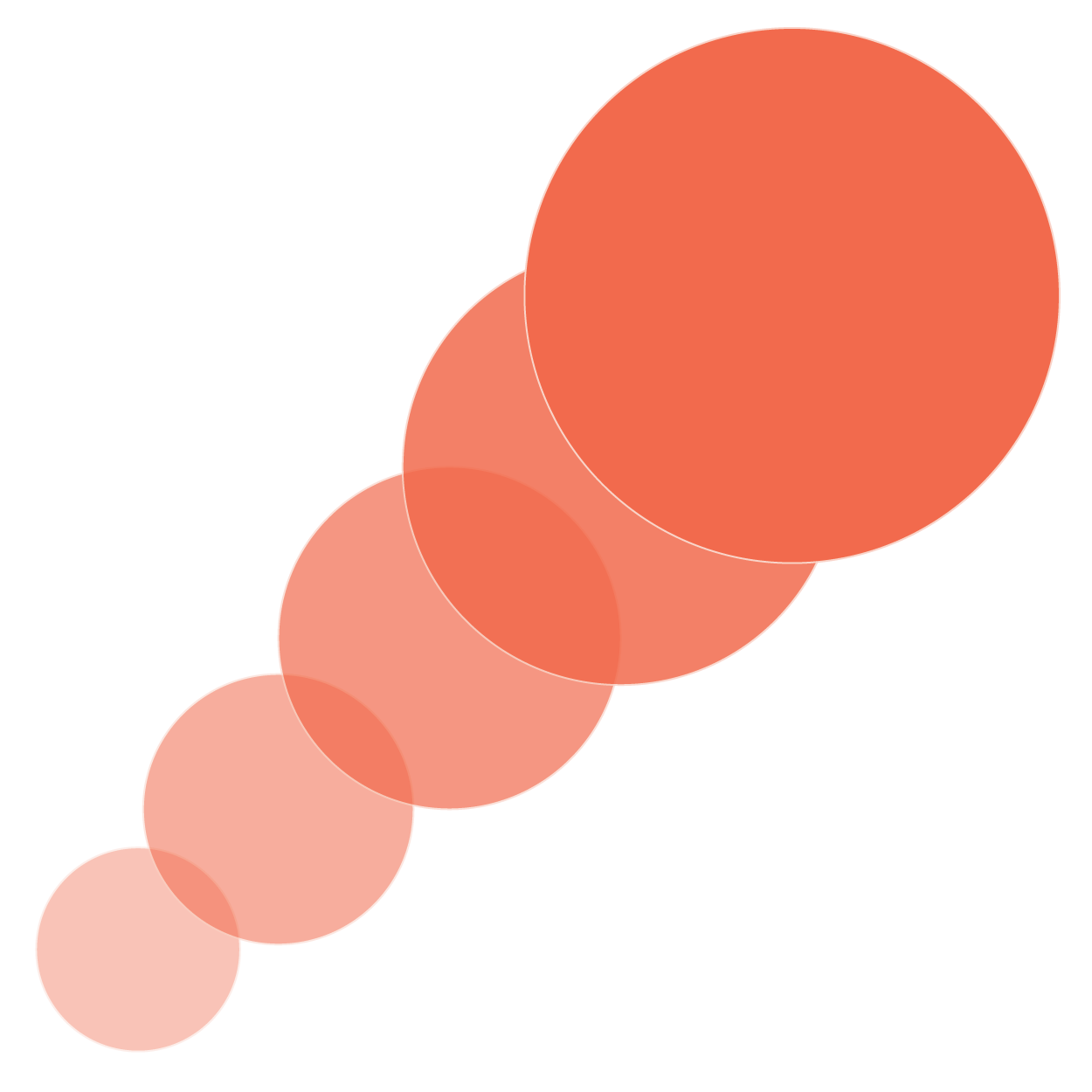}}%
  \endgroup
}
\DeclareRobustCommand{\animated}{\animatedicon\hspace{\iconspace}~\approachbg{Animated}}

\DeclareRobustCommand{\transformationicon}{%
  \begingroup\normalfont
\raisebox{\iconverticaloffset}[0pt][0pt]%
{\includegraphics[height=\iconheight]{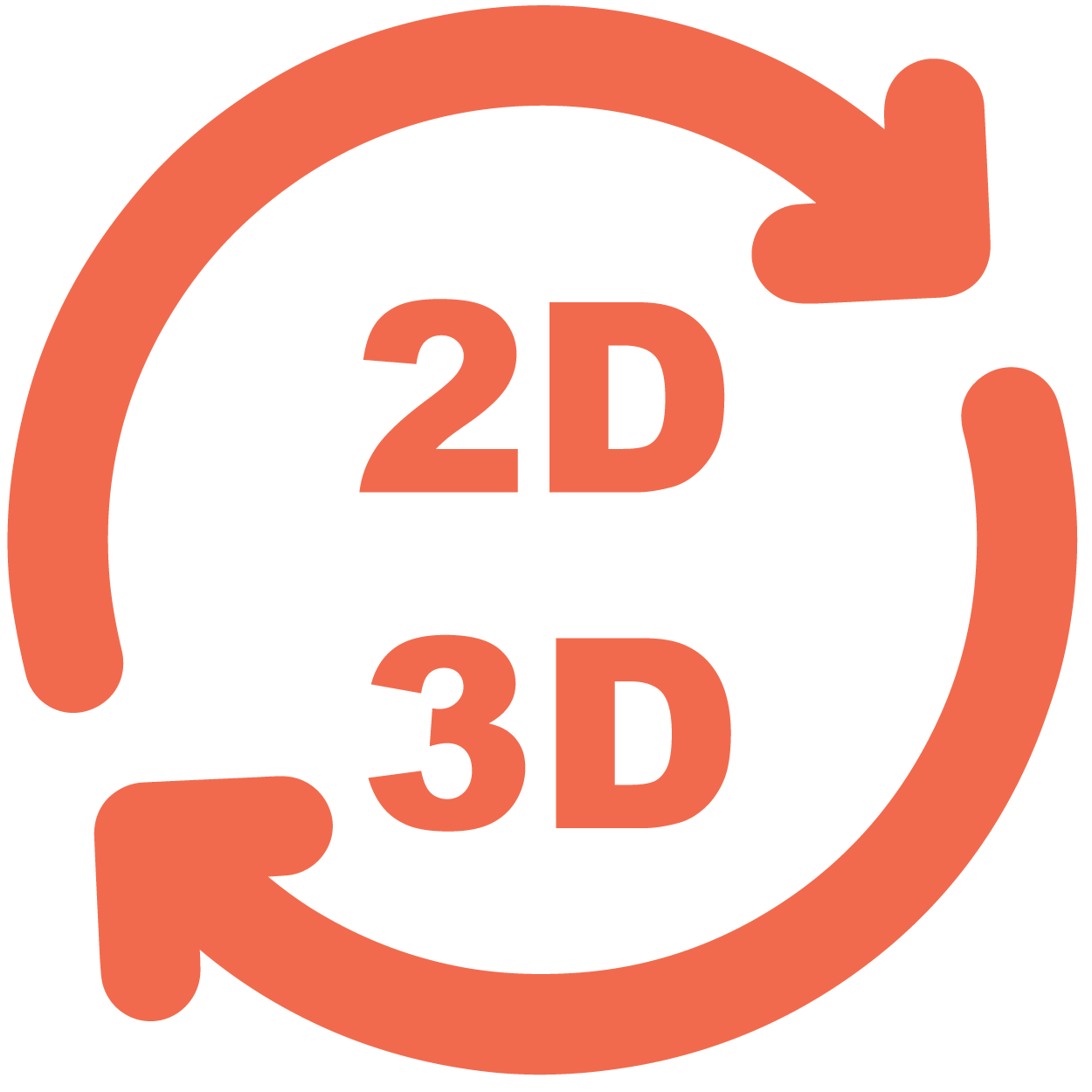}}%
  \endgroup
}
\DeclareRobustCommand{\transformation}{\animatedicon~\transformationicon\hspace{\iconspace}~\approachbg{Transformation}}

\DeclareRobustCommand{\modificationicon}{%
  \begingroup\normalfont
\raisebox{\iconverticaloffset}[0pt][0pt]%
{\includegraphics[height=\iconheight]{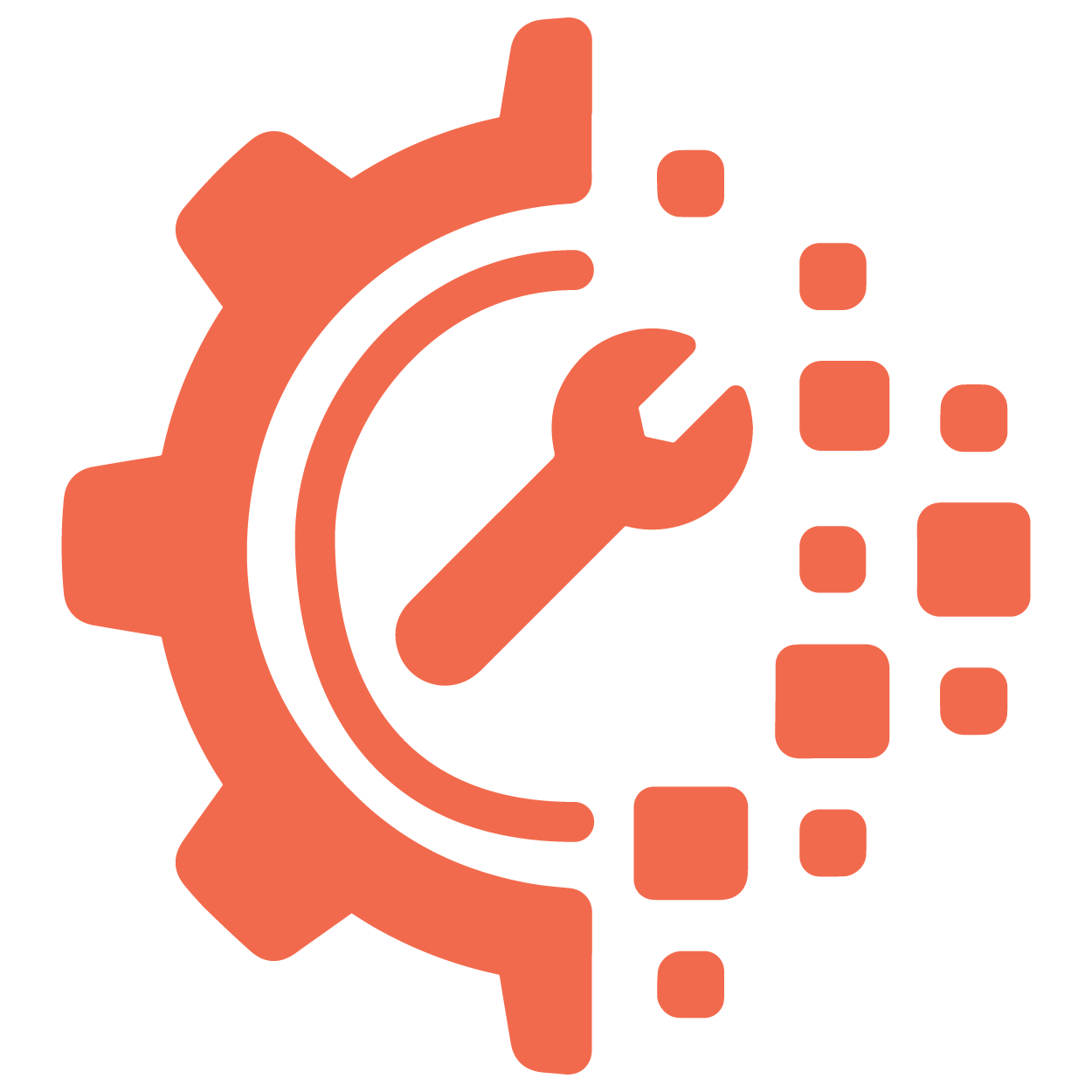}}%
  \endgroup
}
\DeclareRobustCommand{\modification}{\animatedicon~\modificationicon\hspace{\iconspace}~\approachbg{Modification}}

\DeclareRobustCommand{\juxtaicon}{%
  \begingroup\normalfont
\raisebox{\iconverticaloffset}[0pt][0pt]%
{\includegraphics[height=\iconheight]{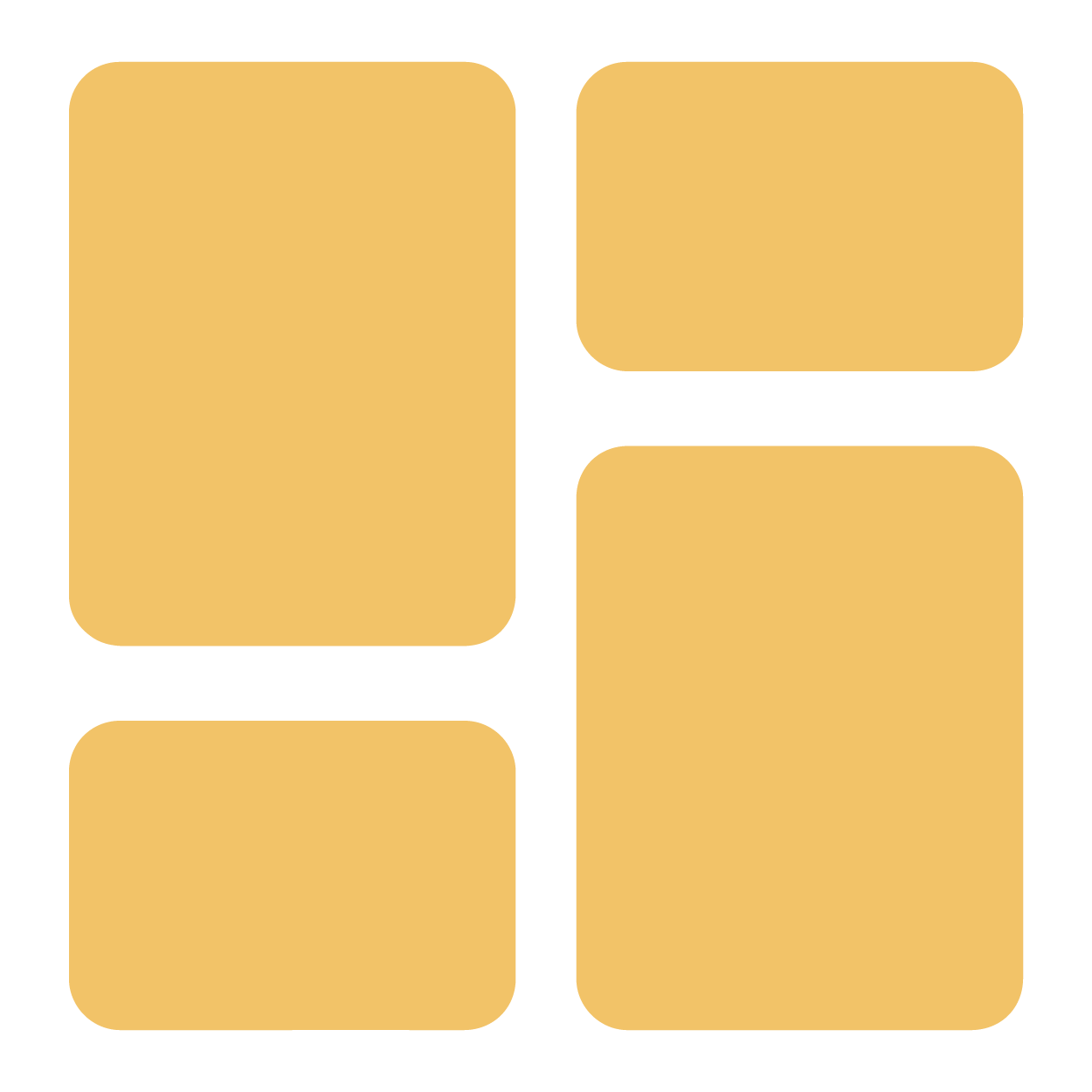}}%
  \endgroup
}
\DeclareRobustCommand{\juxtaposed}{\juxtaicon\hspace{\iconspace}~\layoutbg{Juxtaposed}}

\DeclareRobustCommand{\compoundedicon}{%
  \begingroup\normalfont
\raisebox{\iconverticaloffset}[0pt][0pt]%
{\includegraphics[height=\iconheight]{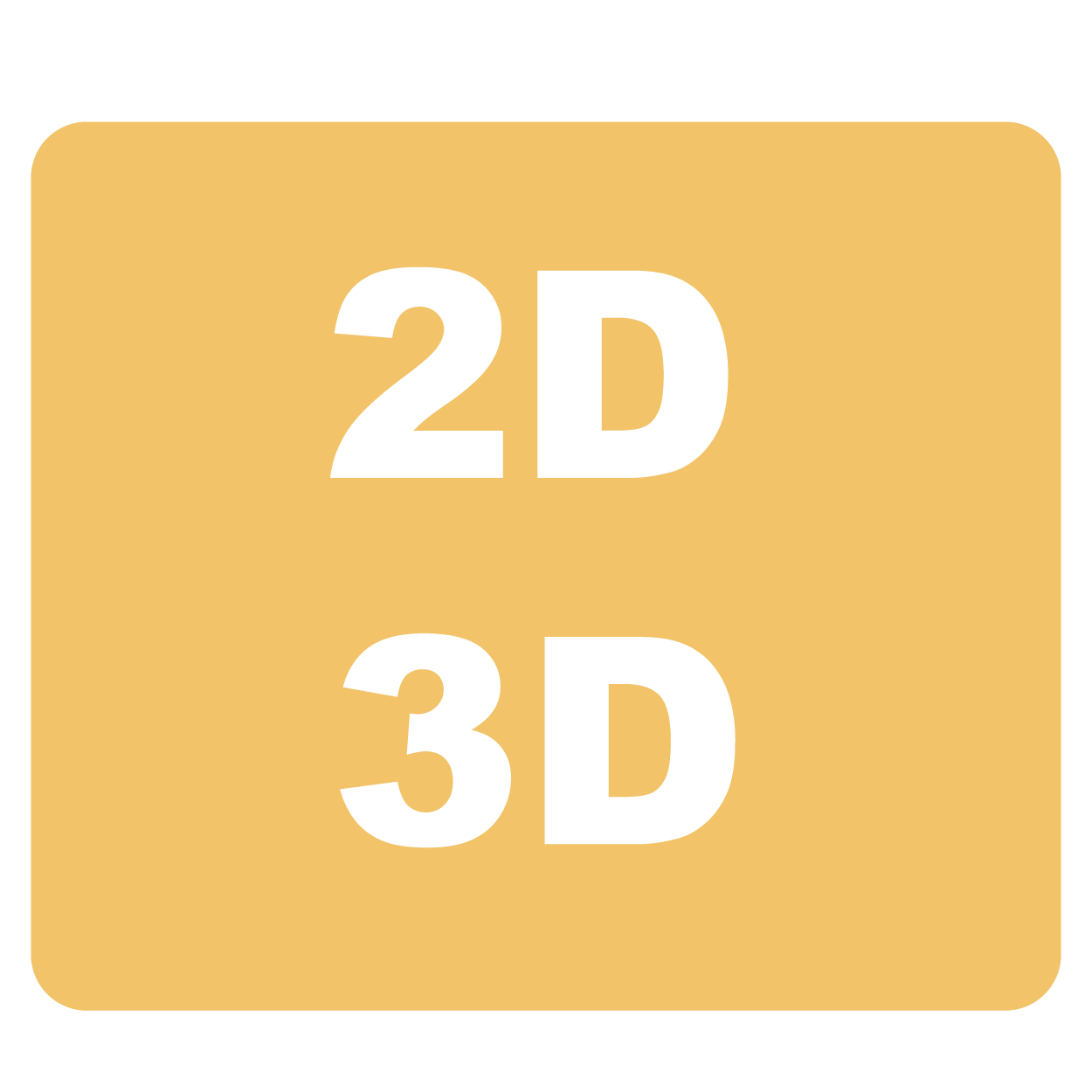}}%
  \endgroup
}
\DeclareRobustCommand{\compounded}{\compoundedicon\hspace{\iconspace}~\layoutbg{Compounded}}

\DeclareRobustCommand{\substituteicon}{%
  \begingroup\normalfont
\raisebox{\iconverticaloffset}[0pt][0pt]%
{\includegraphics[height=\iconheight]{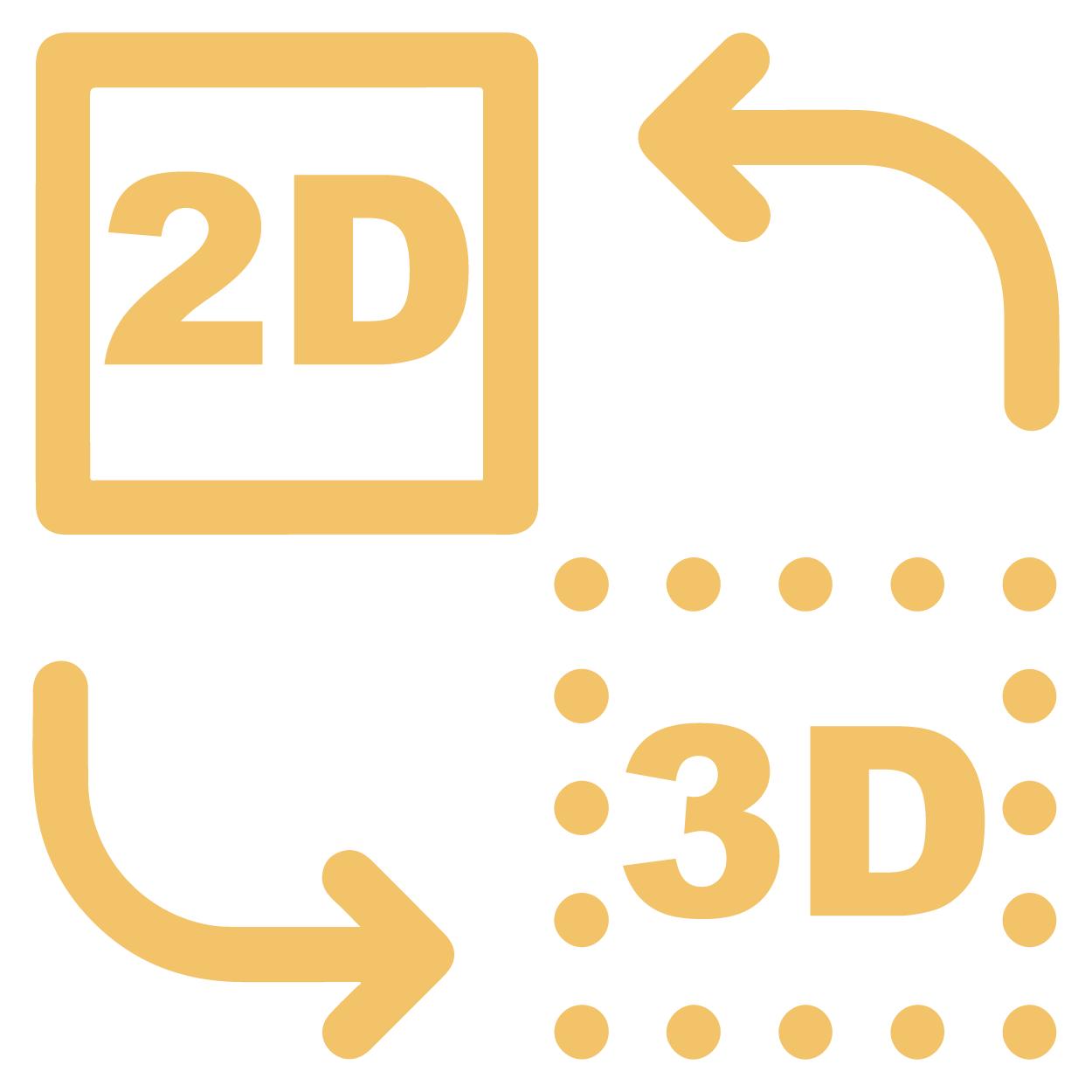}}%
  \endgroup
}
\DeclareRobustCommand{\substitute}{\compoundedicon~\substituteicon\hspace{\iconspace}~\layoutbg{Substituted}}

\DeclareRobustCommand{\superimposedicon}{%
  \begingroup\normalfont
\raisebox{\iconverticaloffset}[0pt][0pt]%
{\includegraphics[height=\iconheight]{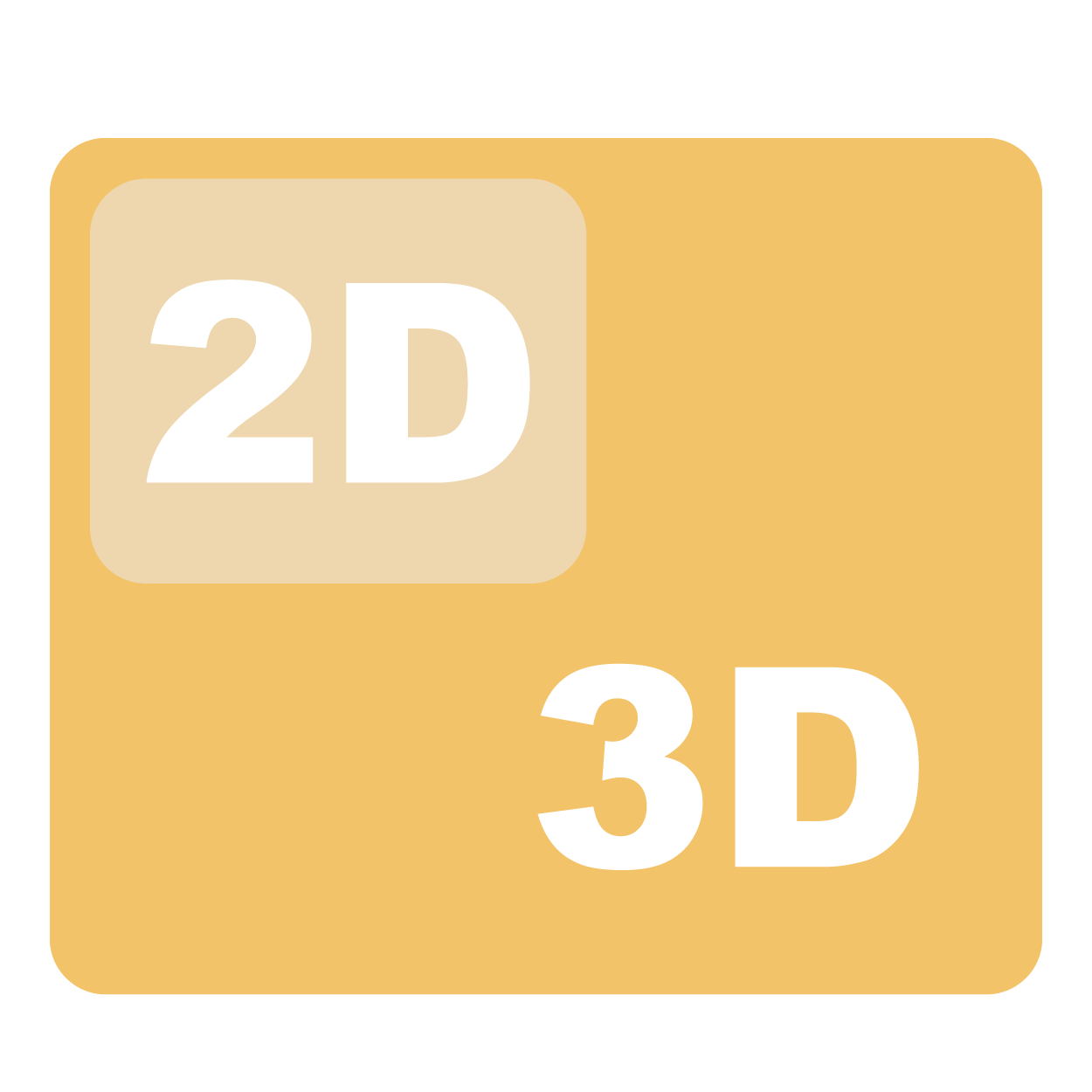}}%
  \endgroup
}
\DeclareRobustCommand{\superimposed}{\compoundedicon~\superimposedicon\hspace{\iconspace}~\layoutbg{Superimposed}}

\DeclareRobustCommand{\embeddedicon}{%
  \begingroup\normalfont
\raisebox{\iconverticaloffset}[0pt][0pt]%
{\includegraphics[height=\iconheight]{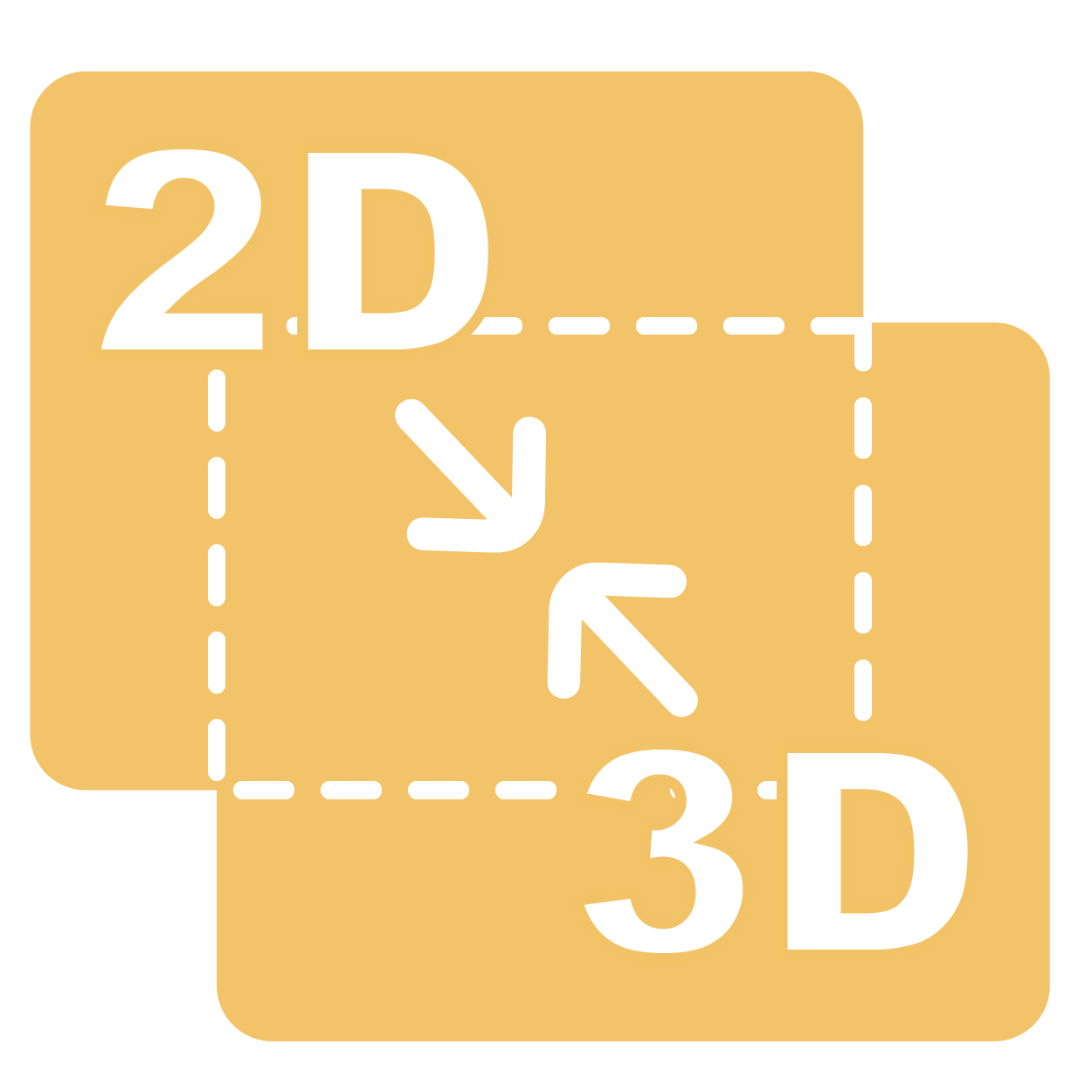}}%
  \endgroup
}
\DeclareRobustCommand{\embedded}{\compoundedicon~\embeddedicon\hspace{\iconspace}~\layoutbg{Embedded}}

\DeclareRobustCommand{\suppleicon}{%
  \begingroup\normalfont
\raisebox{\iconverticaloffset}[0pt][0pt]{\includegraphics[height=\iconheight]{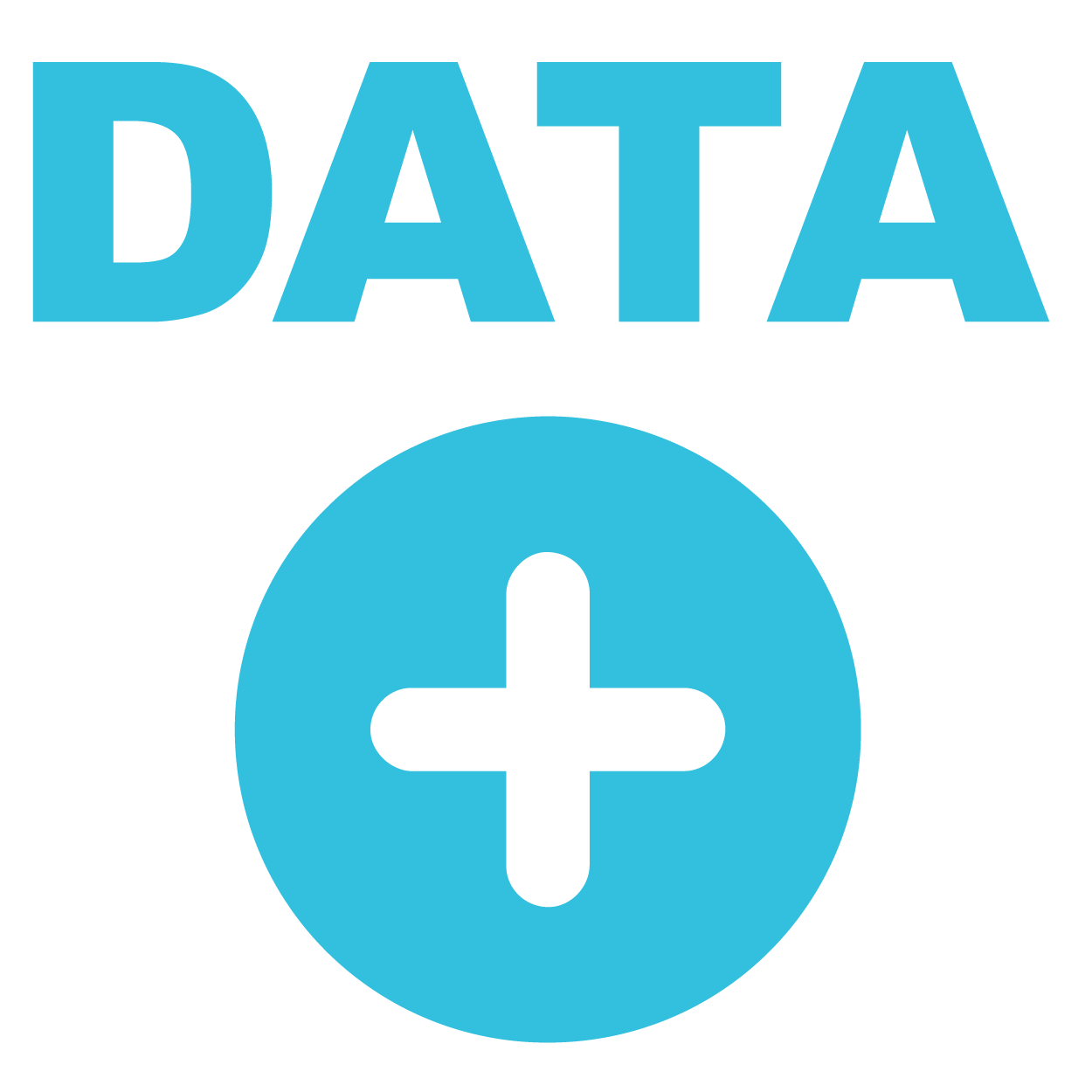}}%
  \endgroup
}
\DeclareRobustCommand{\supplementation}{\suppleicon\hspace{\iconspace}~\motivationbg{Supplementation}}

\DeclareRobustCommand{\abstracticon}{%
  \begingroup\normalfont
\raisebox{\iconverticaloffset}[0pt][0pt]{\includegraphics[height=\iconheight]{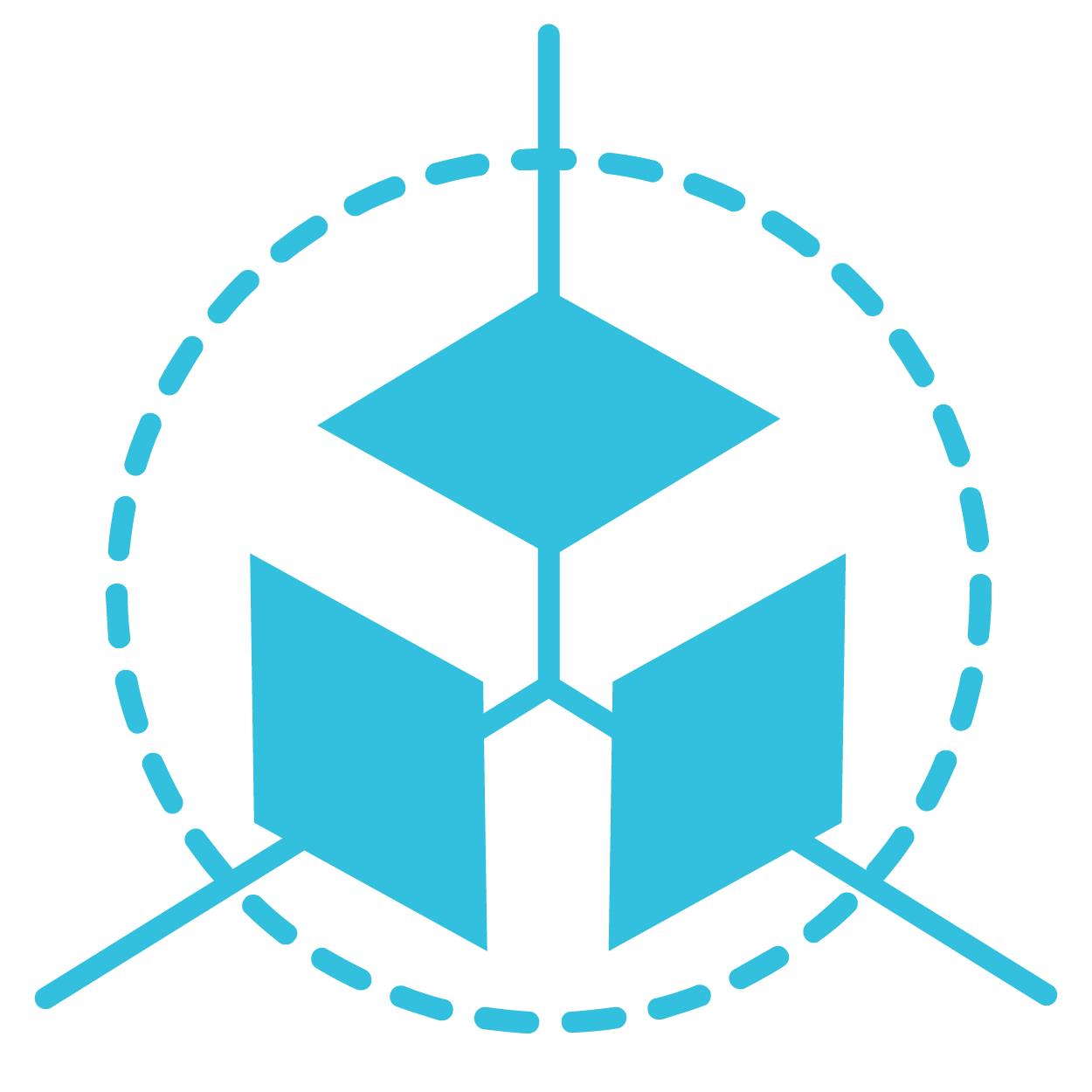}}%
  \endgroup
}
\DeclareRobustCommand{\abstraction}{\abstracticon\hspace{\iconspace}~\motivationbg{Abstraction}}

\DeclareRobustCommand{\projectionicon}{%
  \begingroup\normalfont
\raisebox{\iconverticaloffset}[0pt][0pt]%
{\includegraphics[height=\iconheight]{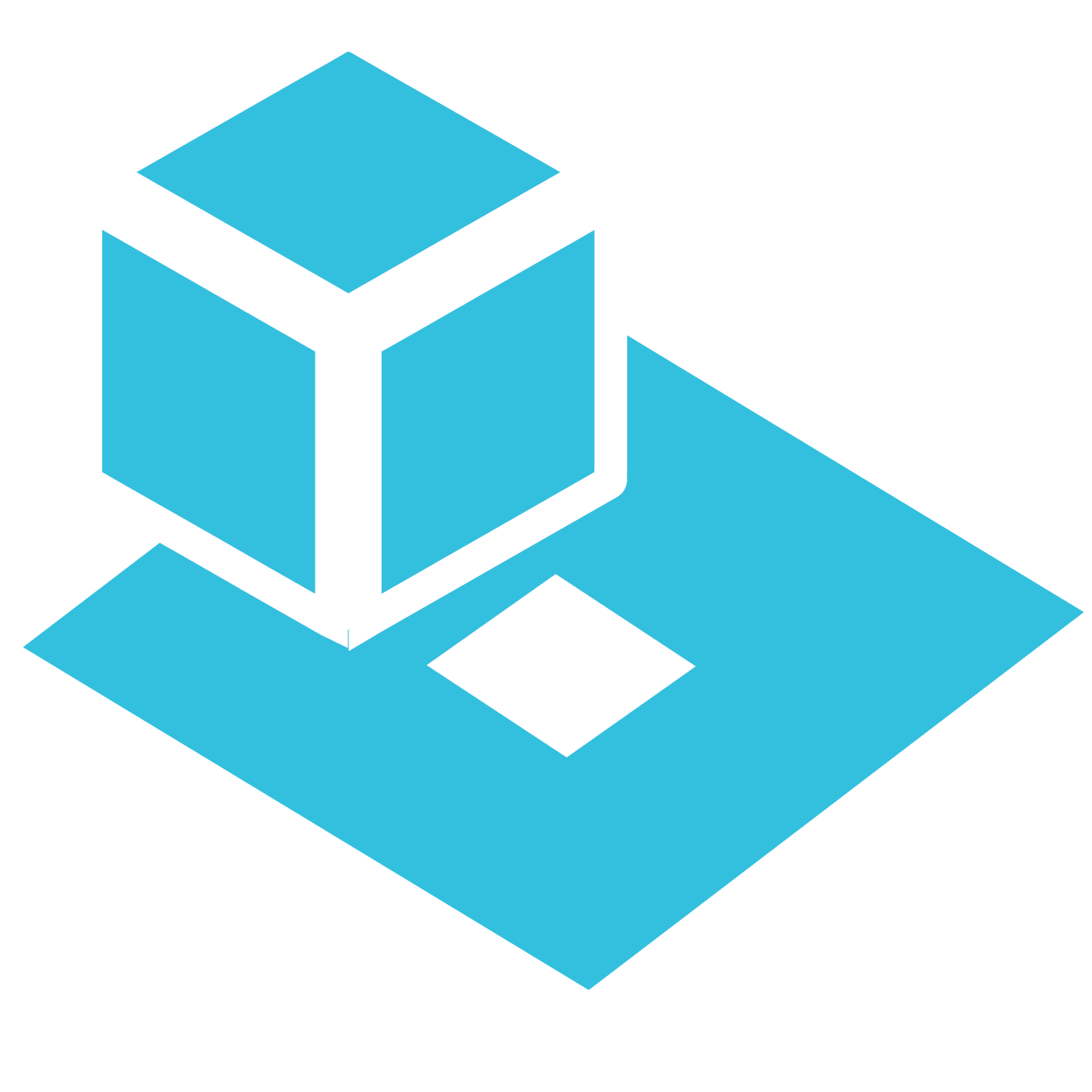}}%
  \endgroup
}
\DeclareRobustCommand{\projection}{\abstracticon~\projectionicon\hspace{\iconspace}~\motivationbg{Projection}}

\DeclareRobustCommand{\flattenicon}{%
  \begingroup\normalfont
\raisebox{\iconverticaloffset}[0pt][0pt]%
{\includegraphics[height=\iconheight]{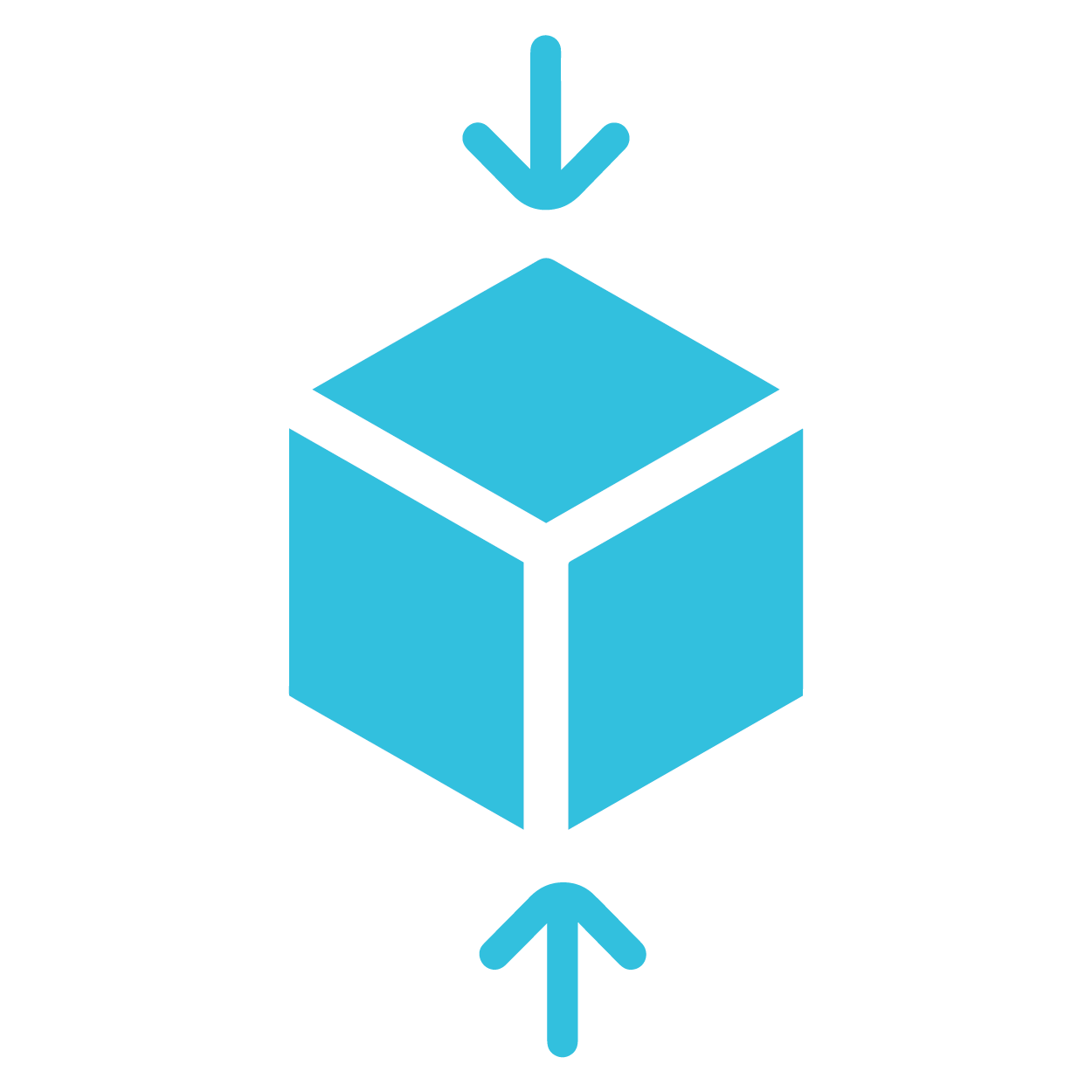}}%
  \endgroup
}
\DeclareRobustCommand{\flatten}{\abstracticon~\flattenicon\hspace{\iconspace}~\motivationbg{Flattening}}

\DeclareRobustCommand{\slicingicon}{%
  \begingroup\normalfont
\raisebox{\iconverticaloffset}[0pt][0pt]%
{\includegraphics[height=\iconheight]{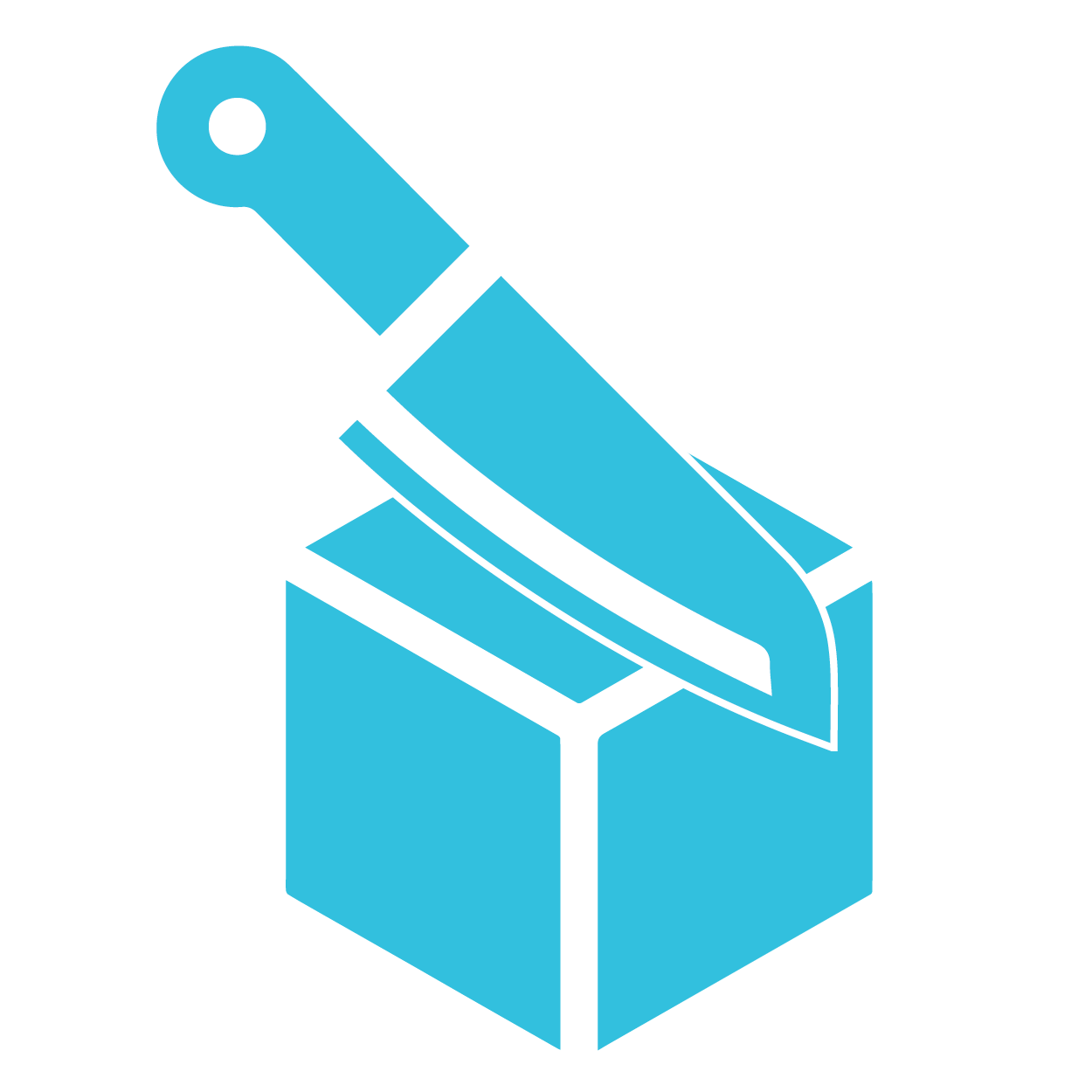}}%
  \endgroup
}
\DeclareRobustCommand{\slicing}{\abstracticon~\slicingicon\hspace{\iconspace}~\motivationbg{Slicing}}

\DeclareRobustCommand{\controlicon}{%
  \begingroup\normalfont
\raisebox{\iconverticaloffset}[0pt][0pt]%
{\includegraphics[height=\iconheight]{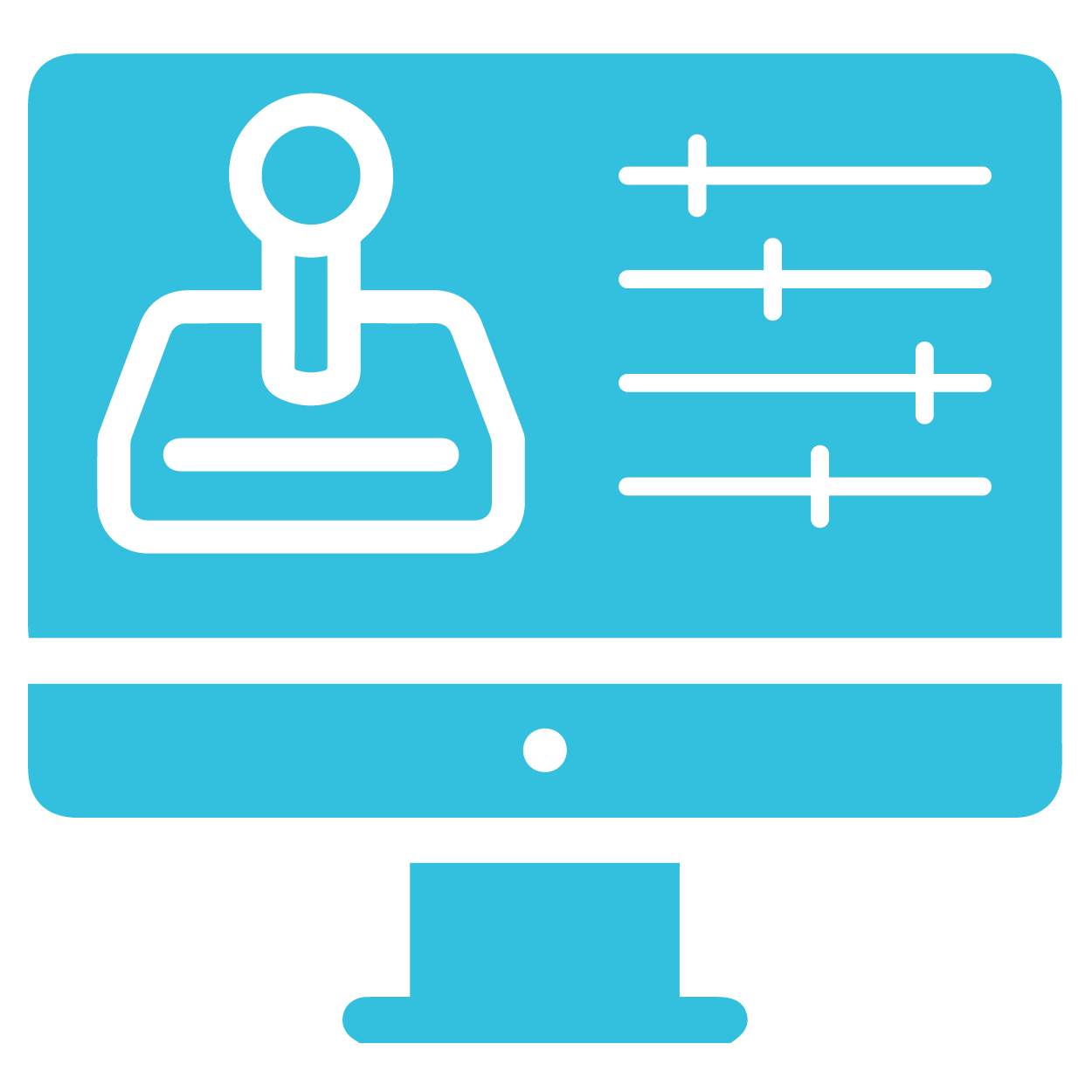}}%
  \endgroup
}
\DeclareRobustCommand{\control}{\controlicon\hspace{\iconspace}~\motivationbg{Control}}

\begin{document}

%%%%%%%%%%%%%%%%%%%%%%%%%%%%%%%%%%%%%%%%%%%%%%%%%%%%%%%%%%%%%%%%
%%%%%%%%%%%%%%%%%%%%%% START OF THE PAPER %%%%%%%%%%%%%%%%%%%%%%
%%%%%%%%%%%%%%%%%%%%%%%%%%%%%%%%%%%%%%%%%%%%%%%%%%%%%%%%%%%%%%%%

\firstsection{Introduction}
\maketitle
% linking
Visual representations are often connected to help viewers locate objects of interest that are displayed in multiple diagrams \cite{Wills:2008:LDV}. Connected representations allow us to examine data from multiple perspectives. This linking can be realized using various means: through interactions, \eg, brushing~\cite{Becker:1987:BS}, focus and linking~\cite{Buja:1991:IDV}, focus+context~\cite{Cockburn:2009:ROD}, and using the same visual encoding~\cite{eulzer:2020:temporal, meuschke:2017:combined}. 
%
% linking 2d and 3d
Our work focuses on semantically-connected 2D and 3D visualizations, \ie, visual representations that are related because they provide information about the same entity.  Linking 2D and 3D visualizations facilitates exploring various levels of detail and abstraction, as well as completing tasks \cite{Tory:2006:VTP}. Rendering objects in 3D is a common technique used in a variety of applications \cite{elvins:1992:survey}, \jh{such as representing the organ structures~\cite{eulzer:2021:visualization} or the universe~\cite{Fu:2007:TSV}}. Unlike their 2D counterparts, 3D representations \jh{use an} additional dimension to encode values \cite{brath:2015:3d} and provide intuitive insight into spatial~\cite{miao:2018:dimsum} or spatiotemporal data~\cite{tominski:2012:stacking}. \jh{With 3D visualizations it can be difficult, however, to navigate and interpret hidden information such as the relationships between 3D objects~\cite{Hong:2021:DET}}. As such, 2D representations are far more common; they are used for visualizing abstract information \cite{mohammed:2018:abstractocyte, smit:2017:pelvis} and other descriptive data \cite{hadwiger:2012:interactive, kottravel:2019:visual}. Previous work has demonstrated that 2D and 3D representations can be more effective when combined with one another~\cite{Tory:2006:VTP, Amini:2015:IIC}, and there is an increasing body of work that combines 2D and 3D visualizations~\cite{tominski:2012:stacking, reh:2013:mobjects, meuschke:2017:combined}. 

% why we wrote this survey
Categorizations currently exist for providing links between multiple views \cite{sun:2022:towards} and transforming 2D and 3D representations in \jh{mixed reality (MR)}~\cite{lee:2022:design}; we found, however, that combinations of 2D and 3D representations have relationships that do not fall into the frameworks presented by these \jh{surveys and design spaces}. For instance, the relationship between two representations when they are embedded in one another~\cite{rocha:2018:illustrative, tominski:2012:stacking, halladjian:2020:scale} cannot be described by \jh{the frameworks because} the representations share the same view and do not transform into each other. \jh{In fact, it seems that existing literature mainly focused on 2D representations shown on traditional displays and 3D representations in MR yet missed the necessity of systematically investigating their combinations in various environments.}

Despite the numerous benefits of linking representations of varying dimensions\jh{\cite{Sedlmair:2009:TBP}}, there has not been a systematic review of methods to effectively classify and design these links. To fill this gap, we surveyed papers from IEEE VIS, EuroVis, and TVCG (only those presented at VIS) from 2012 to 2022 and identified \textbf{\jh{105}} relevant publications. From our survey we created a design space (\autoref{fig:overview_design_space}) with four dimensions to categorize these connections: \emph{\motivationbg{Motivation}} for linking these representations, \emph{\displaybg{Display Environment}}, \emph{\layoutbg{Layout}} of the representations, and \emph{\approachbg{Approaches}} for interacting with the links (see \autoref{fig:overview_design_space}). We also implemented a website for readers to explore this design space. As a whole, our work emphasizes the importance of cognitive connections formed between visual representations of varying dimensions. We aim to assist future designers in building effective links between 2D and 3D visual representations.
In summary, with this work, we contribute:
\begin{itemize}[nosep,left=0pt .. \parindent]
    \item a survey of recent visualization work that links 2D and 3D data representations,
    \item a design space and guidelines for systems with linked 2D+3D data representations, and
    \item an interactive website \jh{(\href{https://2dplus3d.github.io/}{\texttt{2dplus3d\discretionary{}{.}{.}github\discretionary{}{.}{.}io}})}.
\end{itemize}

\section{Background}
\label{sec:background}
We start by discussing the background of 2D and 3D visualizations and their combinations, which motivates our survey. We also base our work on other multiple coordinated view designs.

\subsection{2D and 3D Visual Representations}
\jh{Visual representations use graphical elements to encode data, and the dimensionality of a visualization is determined by the number of dimensions it employs to encode the data.}
2D representations have been broadly explored in the visualization field \cite{ware:2019:information, munzner:2014:visualization} \jh{as they have been shown to effectively encode abstract relationships or 2D spatial data.}
3D representations are largely used in scientific contexts~\cite{Yu:2010:FDI}, \eg, exploring spatial or spatiotemporal data~\cite{byska:2015:molecollar}. 
\jh{They can use unique properties, such as meshes and surfaces, to represent complicated data. They are also more efficient} when people are able to manipulate them \cite{hubona:1997:3d}---due to the fact that 3D representations enhance people's spatial memory \cite{tavanti:2001:2d}. Other than spatial data, 3D representations can also represent trees \cite{pavlopoulos:2008:arena3d, Robertson:1991:CTA} and multidimensional data \cite{peng:2014:extensible}; these representations fall under the category of \textit{3D information visualization} \cite{brath:2015:3d}. 

In our survey, we thus explore the design choices of combining 2D+3D representations. The benefits of linking both types of representations have previously been investigated. Tory et al.~\cite{Tory:2006:VTP} performed experiments to compare the performance of 2D and 3D displays, as well as their combinations. They found that combining displays improves task performance and provides an intuitive experience. Amini et al.~\cite{Amini:2015:IIC} compared 2D and 3D representations of spatiotemporal data and found that they were most effective when combined. In \redsout{these two studies}\jh{this study}, researchers treated the combination as placing 2D and 3D representations separately, but there are also other approaches to combining them beyond positions, \jh{such as data transformation}. Lee et al.~\cite{lee:2022:design}, \eg, explored the design space of transforming 3D into 2D representations in virtual reality. We consider the transformation as one type of 2D+3D combination because they are semantically linked. \jh{Other combination approaches include using an additional operation element to control the statuses of both representations~\cite{Lawonn:2016:OBF}.} Although specific combination approaches have been studied, there is no systematic review of \jh{why, when, and} how to combine 2D+3D representations---which we provide.

% \vspace{-0.5ex}
\subsection{Linking Multiple Visualizations}
\jh{Though 2D and 3D representations are not necessarily always displayed in different views, we can learn from previous work in linking multiple views about their design~\cite{Langner:2018:VCC, sun:2022:towards, hadlak:2015:survey,Meuschke:2023:GGC}.}\redsout{Multiple views are commonly linked in visual systems, and the design of these links has been the subject of much work.}
\jh{For example,} Javed and Elmqvist \cite{Javed:2012:EDS} proposed a design space with guidelines for compositions of multiple visualizations, which they called composite visualization views. Based on this work, Deng et al. \cite{deng:2022:revisiting} explored compositional patterns from visualization papers, and Chen et al. \cite{Chen:2021:CCP} studied effective multiple-view visualization designs. Both of them discussed the multiple visualization layout. Similarly, Bach et al. \cite{bach:2022:dashboard} investigated design patterns for dashboards. Sun et al. \cite{sun:2022:towards} followed by exploring cross-view data visualizations, which culminated in design recommendations for linking multiple views.
Also relevant for our work are techniques that link visualizations between different devices. 
Badam and Elmqvist \cite{Badam:2019:VCV} worked on cross-de\-vice visualization and interaction. Satriadi et al. \cite{satriadi:2020:MAM} explored how multiple views can be used in VR settings. They found that viewers prefer egocentric arrangements, \ie, arranging views around a first-person reference point, which provides a better overview of the system and facilitates detailed exploration. 
All of the aforementioned work discusses 2D representations without 3D. Although another related publication \cite{Liu:2020:DEI} discusses 3D visualizations where Liu et al. compared the layout of multiple 3D representations in VR, most previous work discusses multiple views for 2D visual representations, excluding cases where there are 3D visualizations and different visualizations share the same view. Yet, in linking 2D and 3D visual representations we can still have both types of representations in the same view. These result in the fact that the current surveys in linking multiple visualizations cannot directly be applied to 2D+3D representation combinations---which is a goal we pursue with our work.

\section{Design Space}
Despite the prevalence and utility of combining 2D and 3D representations as we described above, to the best of our knowledge no design space has been established that provides researchers and practitioners with guidance on how to build respective 2D+3D visualization systems. Based on previous surveys on linking multiple visualizations as well as composite visualizations, we extracted a set of design dimensions for which 2D and 3D representations can be linked. This linking is often more complicated than regular linked views, and there are a number of unique approaches that do not get utilized when linking two representations of the same type. We first describe our general approach before we describe the details of our design space.

% \vspace{-0.2em}
\subsection{Method}
\label{sec:method}
We initially collected every paper from IEEE VIS, EuroVis, and TVCG articles presented at VIS from 2012 to 2022 that used both 2D and 3D visual representations. One author then skimmed all papers in the set, took notes, and removed those that did not fit our criteria (\Cref{inclusion_criteria}).
We then discussed the mutual points of all papers from the notes and built a preliminary version of the design space. Then, the author who skimmed the papers tagged each paper with its appropriate design dimensions. After they initially categorized each paper, we discussed the design space again to make sure that its structure was clear and that each category was reasonable. We adjusted the design space if we found that it could not describe a paper that was relevant. 

Next, another author reviewed each paper to verify its categorization.
They carefully read it, watched its accompanying video, and ran its demo if available. If we could not find any relevant material, we e-mailed the authors to clarify their use of 2D+3D techniques in the work. If we did not get a reply from the authors, we used our best judgment to decide whether a paper should be included or not. 
We discussed any disagreement between the two examiners in regular meetings with all authors until we all reached a consensus. Each paper was examined by at least two authors to increase the reliability of the categories. We note that there is other related work that connects 2D and 3D representations; however, these papers are beyond our systematic collection. When relevant, we refer to those papers during the discussion.

% \vspace{-0.2em}
\subsection{Inclusion Criteria}
\label{inclusion_criteria}
We arrived at a set of inclusion criteria through our paper selection process and our study of previous work. We define 2D and 3D representations based on the work of D\"ubel et al.~\cite{Dubel:2014:23P}, who introduced the concepts of attribute space \textit{A} (the rendering of the data), reference space \textit{R} (the data being referenced), and $S^N$ to represent the dimensionality (\textit{N}) of the space (\textit{S}). The number of dimensions used to encode data determines the type of representation for us. We treat $\textit{A}^2$ \(\oplus\) $\textit{R}^2$ as 2D representations. \jh{We excluded sliders and buttons from the category of 2D visualizations, such as done by Marton et al. \cite{Marton:2019:FGE}.} Due to this definition, we consider some abstract 3D data mappings~\cite{brath:2015:3d} as 2D representations, \eg, tilted line charts~\cite{brath:2015:3d}. \jh{As a result, we filtered out papers with 2D representations in 3D environments (\eg, \cite{reipschlager:2020:personal}).} In contrast, we consider $\textit{A}^3$ \(\oplus\) $\textit{R}^3$ as 3D representations. Finally, $\textit{A}^2$ \(\oplus\) $\textit{R}^3$ and $\textit{A}^3$ \(\oplus\) $\textit{R}^2$ are embedded 2D and 3D visual representations. 

It can be difficult to determine the dimensions of representations when data is encoded with physical objects because all of them are in a 3D environment. We kept our criteria as the number of dimensions used to encode data, and some representations are not three-di\-men\-sio\-nal within this inclusion criteria. Zooids \cite{goc:2016:zooids}, \eg, are interactive robots that rest on a table. They physically move around on a tabletop in groups to form shapes that serve as tangible visual representations of the data they are displaying. As the interface is located on a 2D plane, the third dimension is not encoded. Therefore, we excluded the paper that used Zooids to encode data~\cite{goc:2019:dynamic}.

Once we had defined the 2D and 3D representations on which to focus, we removed those papers that did not link 2D and 3D representations. A paper may include multiple visual representations where only some links meet our inclusion criteria, and a paper was included as long as it contained at least one linking method that fit our criteria. In total, we thus included a set of \jh{105} papers in our survey.

% \vspace{-0.2em}
\subsection{Design Space Dimensions}
To develop our design space and based on our paper summaries (\Cref{sec:method}), we aligned the design dimensions with the one for visualization tasks~\cite{schulz:2013:DSV} because the aim of 2D+3D combination is to complete visualization tasks. We organized the design space based on the ``5 W's'': WHY, WHAT, WHERE, WHO, and WHEN, as well as HOW.
In our concept, we focus on answering on WHY, WHERE, and HOW because WHO and WHEN rarely influence the design space and WHAT remains constant (always refers to 2D and 3D representations). 

We first examined WHY people link two visual representations. 2D and 3D representations are combined because they need the other to assist in accomplishing tasks. This lead to a designer's \textit{\motivationbg{Motivation}}, which can vary from providing additional information to providing advanced interactions. Then, with representations and goals established, we investigated WHERE the links are placed, specifically the \textit{\displaybg{Display Environment}}---the input and output devices that display the visual representations. Finally, we explored ways of HOW 2D+3D representations are linked from two aspects: their \textit{\layoutbg{Layout}} and \textit{\approachbg{Approach}}. The \textit{\layoutbg{Layout}} of each representation facilitates mental links, while the \textit{\approachbg{Approach}} emphasizes the relationships of the components within each representation. We discuss the details of each design dimension below.

\begin{table*}[t]
\caption{Reasons of WHY to connect 2D and 3D representations. \jh{See examples in \autoref{fig:motivations}.}% 
}\vspace{-1ex}
\label{tab:reasons}
\tabulinesep=1.5pt
%\extrarowsep=1pt
\begin{tabu}{X[.7,m]X[.7,m]X[3.5,m]}
\toprule 
\multicolumn2{l}{Motivation}   & Papers\\
\midrule
\multicolumn2{l}{\supplementation}       &
\jh{86 papers} \cite{rocha:2018:illustrative, hadwiger:2012:interactive, beyer:2013:connectomeexplorer, sabando:2021:chemva, meuschke:2016:semi, tierny:2018:topology, eulzer:2020:temporal, nguyen:2021:visualization, doraiswamy:2021:topomap, ye:2021:shuttlespace, waser:2014:many, hurter:2019:fiberclay, reh:2013:mobjects, tominski:2012:stacking, meuschke:2017:combined, smit:2017:pelvis, vazquez:2018:visual, byska:2015:molecollar, demir:2014:multi, zhang:2014:visualizing, miao:2018:dimsum, awami:2016:neuroblocks, eulzer:2021:visualizing, zhang:2012:knotpad, agus:2019:interactive, landge:2012:visualizing, kottravel:2019:visual, lindow:2019:interactive, hong:2014:flda, masood:2021:visual, aboulhassan:2015:novel, berge:2014:predicate, biswas:2013:information, burchett:2019:igm, duran:2019:visualization, elek:2021:polyphorm, frohler:2019:visual, furmanova:2020:multiscale, gu:2016:mining, hollt:2012:seivis, ip:2012:hierarchical, krone:2013:interactive, lichtenberg:2018:analyzing, luciani:2019:details, maries:2013:grace, mistelbauer:2013:vessel, palenik:2020:scale, palmas:2014:movexp, poco:2012:employing, rapp:2021:visual, rieck:2012:multivariate, siddiqui:2021:progressive, tao:2019:exploring, unger:2012:visual, wentzel:2020:cohort, zhou:2021:data, bock:2016:visual, cornel:2015:visualization, meuschke:2019:visual, schroeder:2014:trend, weissenbock:2019:dynamic, saffo:2020:data, bader:2020:extraction, ferstl:2017:time, neugebauer:2013:amnivis, hermosilla:2017:physics, awami:2014:neurolines, liu:2021:suggestive, guo:2013:coupled, semmo:2012:interactive, eulzer:2021:visualization, axelsson:2017:dynamic, lawonn:2023:gray, neuroth:2023:level, ulbrich:2023:smolboxes, ageeli:2023:multivariate, splechtna:2023:interactive, nipu:2023:visual, huang:2023:flownl, troidl:2022:barrio, hong:2022:lineaged, yang:2019:origin, kohler:2018:visual, jackson:2013:lightweight, Lawonn:2016:OBF, Klemm:2014:IVA} \\
\hline
\multirow{2}{*}{\abstraction}     
& {\projection}  &  \jh{7 papers} \cite{kolesar:2017:fractional, sabando:2021:chemva, doraiswamy:2021:topomap, halladjian:2020:scale, saffo:2020:data, lawonn:2023:gray, neuroth:2023:level} \\
& {\flatten}  &  \jh{14 papers} \cite{yang:2018:maps, guo:2016:extracting, eulzer:2020:temporal, nadeem:2017:corresponding, meuschke:2017:combined, mohammed:2018:abstractocyte, miao:2018:dimsum,
eulzer:2021:visualizing, lindow:2019:interactive, neugebauer:2013:amnivis, hermosilla:2017:physics, awami:2014:neurolines, liu:2021:suggestive, krone:2017:molecular} \\
& {\slicing}  &  \jh{21 papers} \cite{hadwiger:2012:interactive, kohler:2018:visual, nguyen:2021:visualization, smit:2017:pelvis, byska:2015:molecollar, eulzer:2021:visualizing, abbasloo:2016:visualizing, berge:2014:predicate, elek:2021:polyphorm, glaber:2014:combined, haehn:2014:design, hollt:2012:seivis, jonsson:2016:intuitive, klein:2012:design, kretschmer:2013:interactive, kretschmer:2014:adr, mistelbauer:2013:vessel, schroeder:2014:trend, eulzer:2021:visualization, besancon:2017:hybrid, Lawonn:2016:OBF}
\\
\hline
\multicolumn2{l}{\control}                                                        & \jh{24 papers} \cite{waser:2014:many, hurter:2019:fiberclay, mohammed:2018:abstractocyte, miao:2018:dimsum, hong:2014:flda, abbasloo:2016:visualizing, aboulhassan:2015:novel, berge:2014:predicate, biswas:2013:information, frohler:2019:visual, glaber:2014:combined, haehn:2014:design, kretschmer:2013:interactive, kretschmer:2014:adr, palenik:2020:scale, tao:2019:exploring, neugebauer:2013:amnivis, hermosilla:2017:physics, liu:2021:suggestive, guo:2013:coupled, axelsson:2017:dynamic, ulbrich:2023:smolboxes, huang:2023:flownl, hong:2022:lineaged} \\
\bottomrule
\label{table:motivation}
\end{tabu}
		\vspace{-1em}
\end{table*}

\begin{figure*}
    \centering
    \includegraphics[width=\linewidth]{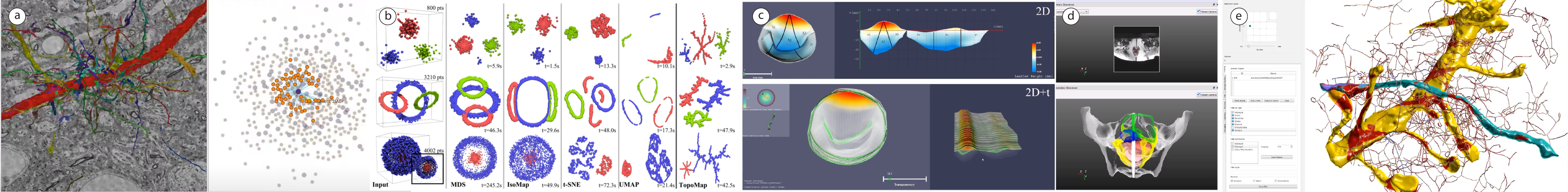}\vspace{-1ex}
    \caption{Examples of various motivations for linking 2D+3D representations (a) \supplementation\ where 2D and 3D representations provide additional information for one another~\cite{beyer:2013:connectomeexplorer}; (b) \jh{\projection\ as it maps 3D data to a 2D plane using the geometry-preserving techniques~\cite{doraiswamy:2021:topomap}}; (c) \flatten\ where 3D mitral valve meshes (bottom) are flattened (top) to facilitate locating regions on the mesh \cite{eulzer:2020:temporal}; \jh{(d) \slicing\ where it renders 2D slices (top) next to 3D volume renders (bottom)~\cite{smit:2017:pelvis};} (e) link providing \control---the 2D diagram on the left controls the cell on the right \cite{mohammed:2018:abstractocyte}.}
    \label{fig:motivations}
\end{figure*}

%\vspace{-0.5em}
\subsection{Motivation: WHY}
When designing a link between 2D and 3D representations, the first factor to consider is the \textit{\motivationbg{Motivation}} as it influences the rest of the design. The reason for connecting 2D and 3D representations often aligns with the purpose of designing visualizations: to facilitate the completion of tasks~\cite{munzner:2014:visualization}. We grouped all the motivations based on these tasks into three categories: \supplementation, \abstraction, and \control---based on how the two types of representations work together to complete tasks (\autoref{fig:motivations}). There are cases where one motivation cannot fully describe the need to have both 3D and 2D representations and we then include multiple motivations. We list all tagged papers in \autoref{table:motivation}.
% \marginpar{\todo{notice the missing ref in \autoref{fig:motivations} (and check that it is indeed an IEEE ref)}}

The first motivation type is \supplementation\ where 2D and 3D representations provide different information for the same object(s).
In this case, completing a task requires information from both representations. 2D representations can present extra data for 3D objects, such as context \cite{bader:2020:extraction} and user interaction history \cite{zhang:2012:knotpad}, while 3D representations can encode other values for 2D diagrams, \eg, temporal \cite{tominski:2012:stacking} and spatial~\cite{kottravel:2019:visual} data. Placing two representations in one interface provides an overview of the problem and facilitates the decision-making process. 

In some situations, people can finish a task with either 2D or 3D representations. In this case, the purpose of combining these representations is to provide information about different aspects of the same object. The 2D and 3D representations share the same data, and the 2D representations often serve as an \abstraction\ of the 3D visualization. This type of relationship usually forms a focus+context view of the target objects. 3D views can also show different visual abstractions~\cite{Viola:2018:PCA, Viola:2020:VA} of spatial data (\eg, \cite{mohammed:2018:abstractocyte}). %TI: removed \cite{zwan:2011:illustrative}

There are \jh{three} types of \abstraction\ techniques employed for generating 2D and 3D views: \projection, \flatten, \jh{and \slicing}. \projection\ \redsout{involves selecting a subset of 3D datasets}\jh{reduces the dimension count} and shows \jh{data typically} in 2D diagrams. \jh{Projections can show, \eg, the inner structure of 3D objects without providing extra information.}
\flatten\ uses \jh{techniques such as parameterization to map 3D objects onto a 2D plane~\cite{eulzer:2020:temporal}. Finally, \slicing\ indicates cases where 2D images are cross-section views of 3D objects or slices from microscopes, \eg, \cite{schroeder:2014:trend, mistelbauer:2013:vessel}. It usually offers additional data}. \jh{For microscope slices,}\redsout{case, 3D representations cannot fully show the detailed structures.} the 2D representations add further context to the 3D representation, so we consider them to be examples of \supplementation. 

The third possible motivation is \control. These are representations included in a system as a means to control other representations, \eg, \cite{mohammed:2018:abstractocyte,miao:2018:dimsum}. \control\ is usually accompanied by another type of motivation because visual representations have data to encode, which means they contain information for \supplementation\ or \abstraction. In this case, \jh{we did not consider those interfaces with operation elements as 2D representations. Normally,} there are one or more primary representations and one or more secondary visualizations, where the secondary representations manipulate the primary visualizations.

%\vspace{-0.5em}
\subsection{Display Environment: WHERE}
\label{design_space:display_environment}
% reminder: details of one paper should only appear once
% detailed example
\begin{table*}[]
\caption{The display environment of each paper we surveyed (WHERE). \jh{See examples in \autoref{fig:environments}.}}\vspace{-0.5ex}
\label{tab:display_environment}
\tabulinesep=1.5pt
%\extrarowsep=1pt
\begin{tabu}{X[1,m]X[4,m]}
\toprule 
Display Environment    
& Papers \\
\midrule
\desktop                            
& \jh{96 papers} \cite{rocha:2018:illustrative, kolesar:2017:fractional, hadwiger:2012:interactive, beyer:2013:connectomeexplorer, guo:2016:extracting, sabando:2021:chemva, meuschke:2016:semi, tierny:2018:topology, kohler:2018:visual, eulzer:2020:temporal, nguyen:2021:visualization, doraiswamy:2021:topomap, nadeem:2017:corresponding, waser:2014:many, reh:2013:mobjects, tominski:2012:stacking, meuschke:2017:combined, smit:2017:pelvis, vazquez:2018:visual, byska:2015:molecollar, halladjian:2020:scale, demir:2014:multi, mohammed:2018:abstractocyte, miao:2018:dimsum, awami:2016:neuroblocks, eulzer:2021:visualizing, zhang:2012:knotpad, agus:2019:interactive, landge:2012:visualizing, kottravel:2019:visual, lindow:2019:interactive, hong:2014:flda, masood:2021:visual, abbasloo:2016:visualizing, aboulhassan:2015:novel, berge:2014:predicate, biswas:2013:information, burchett:2019:igm, duran:2019:visualization, elek:2021:polyphorm, frohler:2019:visual, furmanova:2020:multiscale, glaber:2014:combined, gu:2016:mining, haehn:2014:design, hollt:2012:seivis, ip:2012:hierarchical, kretschmer:2013:interactive, kretschmer:2014:adr, krone:2013:interactive, lichtenberg:2018:analyzing, luciani:2019:details, maries:2013:grace, mistelbauer:2013:vessel, palenik:2020:scale, palmas:2014:movexp, poco:2012:employing, rapp:2021:visual, rieck:2012:multivariate, siddiqui:2021:progressive, tao:2019:exploring, unger:2012:visual, wentzel:2020:cohort, zhou:2021:data, bock:2016:visual, cornel:2015:visualization, meuschke:2019:visual, schroeder:2014:trend, weissenbock:2019:dynamic, saffo:2020:data, bader:2020:extraction, ferstl:2017:time, neugebauer:2013:amnivis, chen:2016:visualization, liu:2016:association, hermosilla:2017:physics, awami:2014:neurolines, liu:2021:suggestive, guo:2013:coupled, semmo:2012:interactive, krone:2017:molecular, eulzer:2021:visualization, axelsson:2017:dynamic, lawonn:2023:gray, neuroth:2023:level, ulbrich:2023:smolboxes, ageeli:2023:multivariate, splechtna:2023:interactive, nipu:2023:visual, huang:2023:flownl, troidl:2022:barrio, hong:2022:lineaged, zhang:2014:visualizing, jackson:2013:lightweight, Lawonn:2016:OBF, Klemm:2014:IVA}\\
\hline
\mr                                          
& 4 papers \cite{yang:2018:maps, ye:2021:shuttlespace, hurter:2019:fiberclay, yang:2019:origin}\\
\hline
\touch    
& 3 papers \cite{jonsson:2016:intuitive, klein:2012:design, besancon:2017:hybrid}\\
\hline
\tangible  
& 2 papers \cite{jackson:2013:lightweight, huron:2014:constructing}\\
\bottomrule
\label{table:display_environment}
%\vspace{-1mm}
\end{tabu}
		%\vspace{-1em}
		\vspace{-1em}\vspace{-.5ex}
\end{table*}

\begin{figure*}
    \centering
    \includegraphics[width=\linewidth]{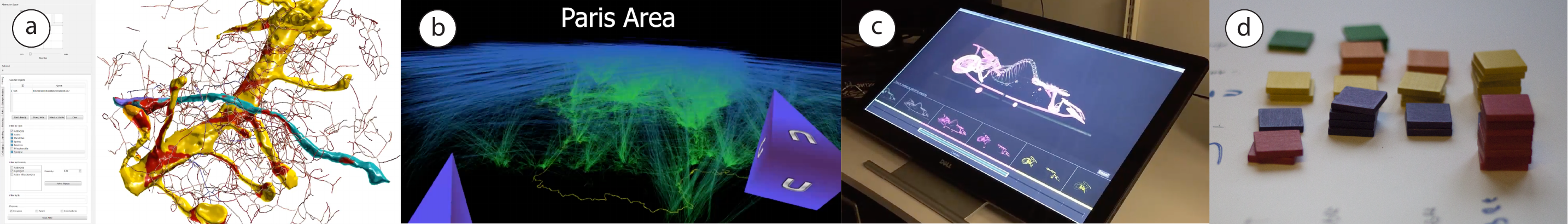}\vspace{-1ex}
    \caption{Examples for different environments: (a) Abstractocyte \cite{mohammed:2018:abstractocyte} as a traditional \desktop\ application; (b) use of \mr\ \cite{hurter:2019:fiberclay}; (c) \touch\ \cite{jonsson:2016:intuitive}; (d) \tangible\ \cite{huron:2014:constructing} showing how vertically combined 2D+3D tangible objects.}\vspace{-.5ex}
    \label{fig:environments}
    %\vspace{-4mm}
\end{figure*}
% \vspace{-0.5em}

With the \textit{\motivationbg{Motivation}} in mind, the next factor to consider when designing a link is its \textit{\displaybg{Display Environment}} (\autoref{tab:display_environment}). These categories are distinguished by their input and output devices. The display environment is important because it can affect the interactions between the representations (\autoref{fig:environments}). We tagged systems with multiple display environments by marking all that applied (as shown in \autoref{table:display_environment}). 

The \desktop\ is the predominant \textit{\displaybg{Display Environment}}. Input is provided through the use of mice and keyboards, and output is displayed on computer monitors. This type of input and output is most familiar to audiences of varying expertise. As such, it is especially useful when performing complicated tasks. Another advantage of desktop environments is the relative maturity of the platform; there are many libraries and tools for building visualizations on desktops. Screen real estate, however, is a major constraint on desktop applications. In addition, because 3D visual representations need to be displayed on a 2D desktop, people may lose some depth cues and spatial understanding~\cite{bowman:2007:virtual}.

Another type of \textit{\displaybg{Display Environment}} is \mr, where both 2D and 3D representations are embedded in immersive environments. Input devices are controllers or gestures, while output devices are typically head-mounted equipment. Immersive systems have been growing in popularity due to their natural ability to convey size and scale in spatial datasets \cite{fonnet:2021:survey}. In addition, virtual reality does not suffer from screen real estate constraints, facilitating the detailed exploration of various views~\cite{hurter:2019:fiberclay}. Complex operations, however, may be difficult to become familiar with due to the nature of the input devices. 
\mr, as a special environment, displays both 2D and 3D in an immersive environment~\cite{marriott:2018:IA}, where 2D representations are typically displayed on a fixed plane \cite{ye:2021:shuttlespace} or rendered in a distorted manner for curved 3D displays \cite{reipschlager:2020:personal}. Theoretically, 2D visual representations can also be placed on an additional device, such as a digital screen~\cite{Langner:2021:MCM}; however, there is no such example in our selections.

Another \textit{\displaybg{Display Environment}} we identified is the \touch. Although the output is often displayed on a digital screen, the input is provided by a person's fingers or hands. This approach achieves a high level of directness in the manipulation space~\cite{Bruckner:2019:MSD}. Haptic interaction methods have been shown to engage people more than traditional input devices~\cite{Hayward:2004:HID}. There are cases in which the \touch\ input device and the output device are not necessarily the same, \eg, \cite{besancon:2017:hybrid}. We do not consider such works to be a \desktop\ application because the gesture can influence the \interactive\ design (\Cref{sec:approach}).

\tangibleP\ are the final set of display environments we examined. With tangible interfaces, physical objects are used to control or form visualizations. As with \mr, 2D representations are displayed in 3D space in this category if there is no other display environment.
When physical objects act as input devices, they are considered to be \emph{tangible user interfaces}~\cite{Ullmer:2000:EFT}. We identified one paper that fit into this group where Jackson et al. \cite{jackson:2013:lightweight} proposed a tangible paper prop as an input device for controlling the 3D orientations of a cross-sectional view of a fiber in virtual reality. 
Physical objects can also represent data in a process termed \emph{data physicalization}~\cite{Jansen:2015:OCD}. Physicalization provides numerous benefits, such as enhancing sensory perception through touch~\cite{Lederman:1987:HMW} and improving the accessibility of data~\cite{Jansen:2015:OCD}. There is one paper in our collection falling into this category. Huron et al.~\cite{huron:2014:constructing} studied the use of tangible items as glyphs. Participants were tasked with creating visual representations using colorful tokens. Placing tokens evenly on a plane created a 2D representation, while stacking them vertically generated a 3D representation (\autoref{fig:environments}(d)). The authors designed three steps for the study and discovered that participants would switch between 2D and 3D representations. They would move and organize the tokens based on their needs, connecting the two types of representations through manual animation.

\vspace{-0.5ex}
\subsection{Layout \& Approach: HOW}
After we have described WHY and WHERE 2D and 3D visualizations are linked, we now investigate HOW they are linked. Two design dimensions exist that link two representations together: \emph{\layoutbg{Layout}} and \emph{\approachbg{Approach}}. The relative layout of each representation provides an inherent emphasis on the relationship between them. The approach directly highlights the relationship between each representation.

%\vspace{-0.5em}
\subsubsection{Layout}
\label{sec:display}
\begin{figure*}
    \centering
    \includegraphics[width=\linewidth]{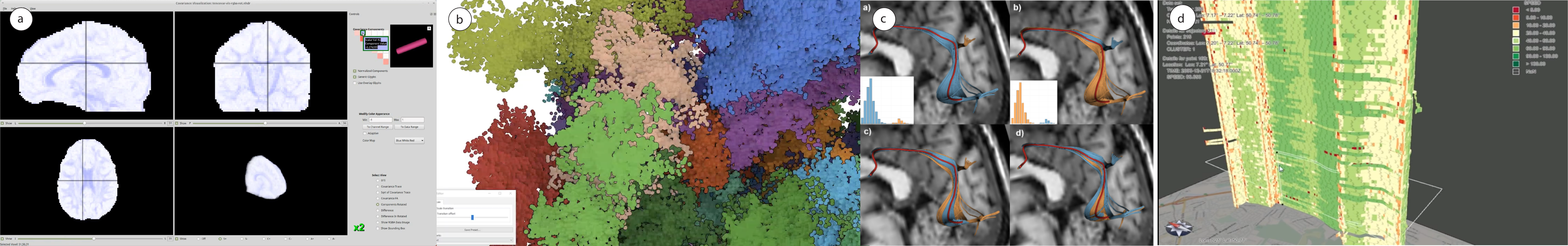}\vspace{-1ex}
    \caption{Examples of layout types: (a) \juxtaposed\ \cite{abbasloo:2016:visualizing}---cross section views are displayed in top-left, top-right, and bottom-left, while the 3D representation is shown in the bottom-right; (b) \substitute\ \cite{halladjian:2020:scale}---the 3D DNA structure can be transformed into 2D representations in one view; (c) \superimposed\ \cite{siddiqui:2021:progressive}---representative fibers are shown in the main 3D view, while the 2D views of fiber confidence intervals are in the bottom-left corner; (d) \embedded\ \cite{tominski:2012:stacking}---the 3D stacked representation shows spatiotemporal data, which is displayed on top of a 2D map.}%\vspace{-1ex}
    \label{fig:layout}
\end{figure*}

\jh{\layoutbg{Layouts} can give designers valuable examples to follow or to be inspired by.} There are two main layout types (\autoref{tab:layout}): \juxtaposed\ and \compounded\ (\autoref{fig:layout}), inspired by composite visualization types~\cite{Javed:2012:EDS}. 

\begin{table*}[]
\caption{The layout of each paper we surveyed (HOW). \jh{See examples in \autoref{fig:layout}.}}\vspace{-0.5ex}
\label{tab:layout}
\tabulinesep=1.5pt
%\extrarowsep=1pt
\begin{tabu}{X[0.4,m]X[0.5,m]X[2,m]}
\toprule 
\multicolumn2{l}{Layout}
& Papers \\
\midrule
\multicolumn2{l}{\juxtaposed} 
& \jh{83 papers} \cite{hadwiger:2012:interactive, beyer:2013:connectomeexplorer, guo:2016:extracting, sabando:2021:chemva, meuschke:2016:semi, tierny:2018:topology, eulzer:2020:temporal, doraiswamy:2021:topomap, nadeem:2017:corresponding, waser:2014:many, hurter:2019:fiberclay, meuschke:2017:combined, smit:2017:pelvis, byska:2015:molecollar, demir:2014:multi, mohammed:2018:abstractocyte, miao:2018:dimsum, awami:2016:neuroblocks, eulzer:2021:visualizing, zhang:2012:knotpad, agus:2019:interactive, landge:2012:visualizing, kottravel:2019:visual, lindow:2019:interactive, hong:2014:flda, masood:2021:visual, abbasloo:2016:visualizing, aboulhassan:2015:novel, berge:2014:predicate, biswas:2013:information, burchett:2019:igm, duran:2019:visualization, elek:2021:polyphorm, frohler:2019:visual, furmanova:2020:multiscale, glaber:2014:combined, gu:2016:mining, haehn:2014:design, hollt:2012:seivis, ip:2012:hierarchical, ip:2012:hierarchical, jonsson:2016:intuitive, klein:2012:design, kretschmer:2013:interactive, kretschmer:2014:adr, krone:2013:interactive, lichtenberg:2018:analyzing, luciani:2019:details, maries:2013:grace, mistelbauer:2013:vessel, palenik:2020:scale, palmas:2014:movexp, poco:2012:employing, rapp:2021:visual, rieck:2012:multivariate, tao:2019:exploring, unger:2012:visual, wentzel:2020:cohort, zhou:2021:data, bock:2016:visual, cornel:2015:visualization, meuschke:2019:visual, schroeder:2014:trend, weissenbock:2019:dynamic, saffo:2020:data, bader:2020:extraction, ferstl:2017:time, hermosilla:2017:physics, awami:2014:neurolines, liu:2021:suggestive, guo:2013:coupled, eulzer:2021:visualization, besancon:2017:hybrid, lawonn:2023:gray, neuroth:2023:level, ulbrich:2023:smolboxes, ageeli:2023:multivariate, splechtna:2023:interactive, nipu:2023:visual, troidl:2022:barrio, hong:2022:lineaged, jackson:2013:lightweight, Lawonn:2016:OBF}\\
\hline
\multirow{3}{*}{\compounded}  & \substitute   & \jh{9 papers} \cite{yang:2018:maps, kolesar:2017:fractional, hurter:2019:fiberclay, halladjian:2020:scale, demir:2014:multi, mohammed:2018:abstractocyte, miao:2018:dimsum, semmo:2012:interactive, krone:2017:molecular}\\
                      & \superimposed  & \jh{9 papers} \cite{hong:2014:flda, elek:2021:polyphorm, siddiqui:2021:progressive, neugebauer:2013:amnivis, ye:2021:shuttlespace, hermosilla:2017:physics, eulzer:2021:visualization, axelsson:2017:dynamic, huang:2023:flownl}\\
                      & \embedded   & \jh{22 papers} \cite{rocha:2018:illustrative, kohler:2018:visual, nguyen:2021:visualization, hurter:2019:fiberclay, reh:2013:mobjects, tominski:2012:stacking, meuschke:2017:combined, smit:2017:pelvis, vazquez:2018:visual, zhang:2014:visualizing, cornel:2015:visualization, meuschke:2019:visual, schroeder:2014:trend, saffo:2020:data, bader:2020:extraction, ferstl:2017:time, hermosilla:2017:physics, guo:2013:coupled, semmo:2012:interactive, eulzer:2021:visualization, yang:2019:origin, Klemm:2014:IVA}\\
\bottomrule
\label{table:layout}
\end{tabu}
		\vspace{-1em}
\end{table*}

Most systems juxtapose two representations; \ie, they place different views side-by-side. 
\juxtaposed\ representations do not overlap, and they can be on different surfaces. \juxtaposed\ layouts provide a clear view of each representation, most useful in cases where both representations are needed during the decision-making process. Although they use more screen space, these layouts enable designers to use various kinds of \textit{\approachbg{Approaches}}. \compounded\ layouts use the same display space or transition between the views and, in contrast, require designers to design custom links---especially \substitute\ and \embedded\ ones. Links on \juxtaposed\ views can be relocated across different devices with the same links. Since they fail to emphasize the connection through their positions, however, they often require the use of additional linking methods to facilitate building mental connections, such as using the same \colorcn\ or adding \animated\ approaches.

Systems that place 2D and 3D representations within the same \compounded\ view use \substitute, \superimposed, and \embedded\ methods.
In substituted layouts, the representations are rendered in the same position at different times. This conserves screen space but leads to a loss of information when one representation is focused. \superimposed\ layouts designate a role for each representation, one being the \textbf{primary} representation, and one being the \textbf{secondary} representation. They are rendered on top of one another.
In most cases, the secondary representation either remains static or reacts to changes in the position from which the primary representation is being observed. Although this approach can suffer from occlusion problems, \superimposed\ layouts can situate representations over target objects to facilitate detailed exploration.

\embedded\ layouts position two representations in the exact same place and emphasize relationships through position. Both representations are likely to move together as the camera rotates. Embedded layouts, however, may also suffer from occlusion. To mitigate this issue, one representation is usually larger and contains more information, while the others add supplementary information.
Such layering (\eg, \cite{tominski:2012:stacking}) avoids the problems of densely packed visualizations. 

When an interface contains multiple 2D and 3D representations, they may contain multiple layouts; \eg, \cite{nguyen:2021:visualization, hurter:2019:fiberclay}. We tagged systems with multiple layouts by marking any layout that was applied.

%\vspace{-0.5em}
\subsubsection{Approach}
\label{sec:approach}

\begin{figure*}
    \centering
    \includegraphics[width=\linewidth]{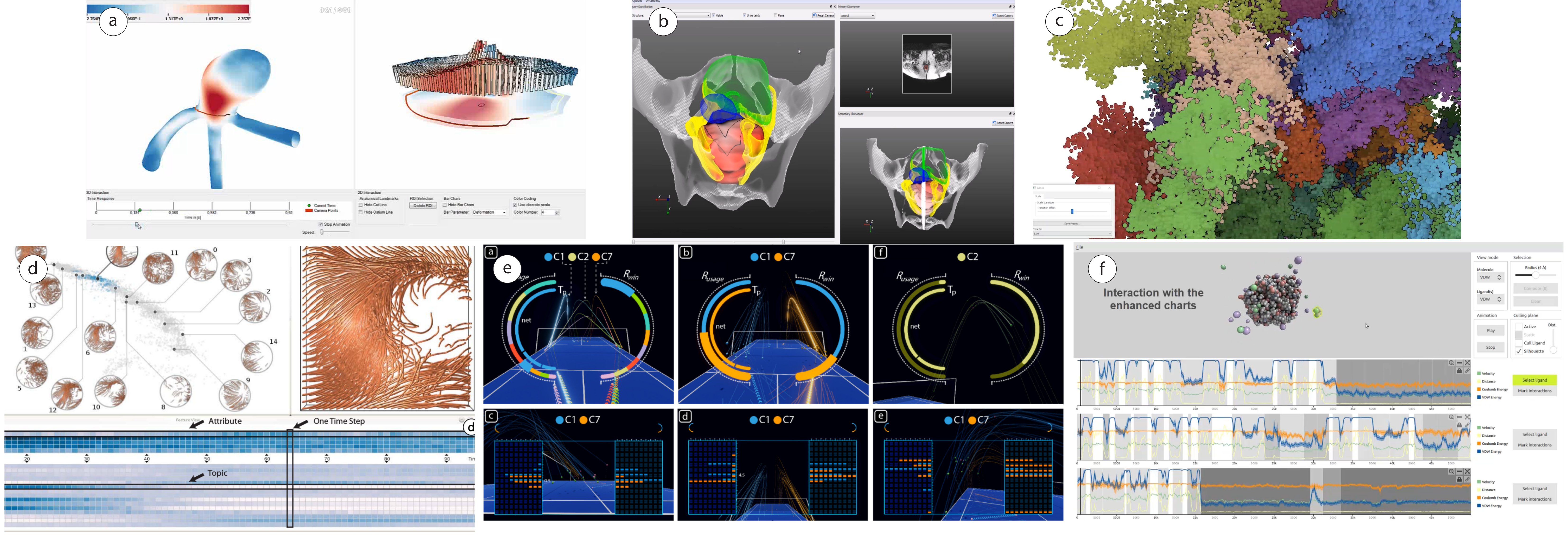}\vspace{-1.5ex}
    \caption{Examples of different ways to connect 2D+3D views. First row---\visual\ methods: \colorcn\ (a)~\cite{meuschke:2017:combined}, \positioncn\ (a)~\cite{meuschke:2017:combined},  \shapecn\ (b)~\cite{smit:2017:pelvis}, and \guidecn\ (d)~\cite{hong:2014:flda}; 
    second row---\interactive\ methods: \twocthree\ (d)~\cite{hong:2014:flda}, \threectwo\ (e)~\cite{ye:2021:shuttlespace}, and \bidirectional\ (f)~\cite{duran:2019:visualization}; third row---\animated\ approaches: \transformation\ (c)~\cite{halladjian:2020:scale} and  \modification\ (a)~\cite{meuschke:2017:combined}.}\vspace{-1.5ex}
    \label{fig:approach}
\end{figure*}

We identified three groups of possible linking approaches: \visual, \interactive, and \animated\ (\autoref{fig:approach}). We listed all the papers in \autoref{table:approach}. \visual\ approaches are cases in which the relationship between the representations is highlighted visually. We found four types of visual encodings to be used: \colorcn, \positioncn, \shapecn, and \guidecn. Visually connecting two representations through \colorcn\ is achieved by using the same color to represent the same object in both representations. Two representations can be visually connected through their \positioncn\ by either placing two representations relative to one another, or having them share their position. If a 3D visualization is embedded in a 2D diagram, \eg, both share their positions. If a cross-section diagram is placed near its corresponding 3D structure, they are linked through their relative positions. Two representations can also be visually connected through \shapecn\ by using the same shapes to represent an object in both representations. \guidecn\ can be used to link visual components between the representations.

Another sub-category within the category of approaches is \interactive\ linking. In this case, viewers interact with one of the representations and get feedback from another. 
We found three sub-types of \interactive\ approach: \twocthree, \threectwo, and \bidirectional. 

\begin{table*}[]
\caption{The linking approaches of selected papers (HOW). \jh{See examples in \autoref{fig:approach}.}}\vspace{-.5ex}
\label{tab:linking_methods}
\tabulinesep=1.5pt
%\extrarowsep=1pt
\begin{tabu}{X[0.9,m]X[0.7,m]X[2.5,m]}
\toprule 
\multicolumn2{l}{Approach}
& Papers \\
\midrule
\multirow{4}{*}{\visual}       & \colorcn                      
                                    & \jh{44 papers} \cite{yang:2018:maps, guo:2016:extracting, meuschke:2016:semi, tierny:2018:topology, eulzer:2020:temporal, nguyen:2021:visualization, nadeem:2017:corresponding, ye:2021:shuttlespace, hurter:2019:fiberclay, meuschke:2017:combined, smit:2017:pelvis, halladjian:2020:scale, demir:2014:multi, mohammed:2018:abstractocyte, eulzer:2021:visualizing, landge:2012:visualizing, lindow:2019:interactive, berge:2014:predicate, biswas:2013:information, frohler:2019:visual, glaber:2014:combined, haehn:2014:design, ip:2012:hierarchical, jonsson:2016:intuitive, klein:2012:design, lichtenberg:2018:analyzing, luciani:2019:details, maries:2013:grace, mistelbauer:2013:vessel, palmas:2014:movexp, poco:2012:employing, rapp:2021:visual, siddiqui:2021:progressive, tao:2019:exploring, unger:2012:visual, zhou:2021:data, meuschke:2019:visual, schroeder:2014:trend, ferstl:2017:time, besancon:2017:hybrid, ulbrich:2023:smolboxes, troidl:2022:barrio, hong:2022:lineaged, Lawonn:2016:OBF}      \\
                                    & \positioncn 
                                    & \jh{30 papers} \cite{rocha:2018:illustrative, yang:2018:maps, kolesar:2017:fractional, sabando:2021:chemva, meuschke:2016:semi, kohler:2018:visual, hurter:2019:fiberclay, reh:2013:mobjects, tominski:2012:stacking, meuschke:2017:combined, vazquez:2018:visual, halladjian:2020:scale, mohammed:2018:abstractocyte, zhang:2012:knotpad, palmas:2014:movexp, poco:2012:employing, cornel:2015:visualization, bader:2020:extraction, ferstl:2017:time, guo:2013:coupled, semmo:2012:interactive, neuroth:2023:level, yang:2019:origin, eulzer:2021:visualization, nguyen:2021:visualization, saffo:2020:data, schroeder:2014:trend, smit:2017:pelvis, zhang:2014:visualizing, Klemm:2014:IVA}      \\
                                    & \shapecn                      
                                    & \jh{35 papers} \cite{yang:2018:maps, kolesar:2017:fractional, hadwiger:2012:interactive, sabando:2021:chemva, kohler:2018:visual, nguyen:2021:visualization, doraiswamy:2021:topomap, smit:2017:pelvis, byska:2015:molecollar, halladjian:2020:scale, mohammed:2018:abstractocyte, eulzer:2021:visualizing, lindow:2019:interactive, abbasloo:2016:visualizing, berge:2014:predicate, elek:2021:polyphorm, glaber:2014:combined, haehn:2014:design, hollt:2012:seivis, jonsson:2016:intuitive, klein:2012:design, kretschmer:2013:interactive, kretschmer:2014:adr, mistelbauer:2013:vessel, schroeder:2014:trend, saffo:2020:data, neugebauer:2013:amnivis, hermosilla:2017:physics, awami:2014:neurolines, liu:2021:suggestive, eulzer:2021:visualization, besancon:2017:hybrid, lawonn:2023:gray, neuroth:2023:level, nipu:2023:visual, Lawonn:2016:OBF}      \\
                                    & \guidecn             
                                    & \jh{5 papers} \cite{doraiswamy:2021:topomap, vazquez:2018:visual, zhang:2014:visualizing, hong:2014:flda, eulzer:2021:visualization}      \\
\hline
\multirow{3}{*}{\interactive}  & \twocthree                   
                                    & \jh{42 papers} \cite{hadwiger:2012:interactive, guo:2016:extracting, sabando:2021:chemva, reh:2013:mobjects, demir:2014:multi, mohammed:2018:abstractocyte, awami:2016:neuroblocks, eulzer:2021:visualizing, agus:2019:interactive, kottravel:2019:visual, hong:2014:flda, abbasloo:2016:visualizing, berge:2014:predicate, biswas:2013:information, frohler:2019:visual, furmanova:2020:multiscale, gu:2016:mining, haehn:2014:design, hollt:2012:seivis, ip:2012:hierarchical, jonsson:2016:intuitive, kretschmer:2014:adr, mistelbauer:2013:vessel, palenik:2020:scale, palmas:2014:movexp, poco:2012:employing, rapp:2021:visual, rieck:2012:multivariate, siddiqui:2021:progressive, tao:2019:exploring, zhou:2021:data, bock:2016:visual, meuschke:2019:visual, schroeder:2014:trend, ferstl:2017:time, awami:2014:neurolines, liu:2021:suggestive, guo:2013:coupled, eulzer:2021:visualization, ageeli:2023:multivariate, nipu:2023:visual, huang:2023:flownl}      \\
                                    & \threectwo                   
                                    & \jh{4 papers} \cite{nadeem:2017:corresponding, ye:2021:shuttlespace, elek:2021:polyphorm, besancon:2017:hybrid}      \\
                                    & \bidirectional          
                                    & \jh{34 papers} \cite{beyer:2013:connectomeexplorer, eulzer:2020:temporal, waser:2014:many, hurter:2019:fiberclay, tominski:2012:stacking, meuschke:2017:combined, smit:2017:pelvis, miao:2018:dimsum, landge:2012:visualizing, lindow:2019:interactive, masood:2021:visual, aboulhassan:2015:novel, burchett:2019:igm, duran:2019:visualization, glaber:2014:combined, klein:2012:design, kretschmer:2013:interactive, lichtenberg:2018:analyzing, luciani:2019:details, unger:2012:visual, wentzel:2020:cohort, cornel:2015:visualization, weissenbock:2019:dynamic, saffo:2020:data, bader:2020:extraction, neugebauer:2013:amnivis, hermosilla:2017:physics, axelsson:2017:dynamic, lawonn:2023:gray, neuroth:2023:level, ulbrich:2023:smolboxes, splechtna:2023:interactive, troidl:2022:barrio, hong:2022:lineaged}      \\
\hline
\multirow{2}{*}{\animated}                & \transformation                  
                                    & \jh{7 papers} \cite{yang:2018:maps, hurter:2019:fiberclay, halladjian:2020:scale, mohammed:2018:abstractocyte, miao:2018:dimsum, semmo:2012:interactive, krone:2017:molecular}      \\
                                    & \modification   
                                    & \jh{7 papers} \cite{meuschke:2017:combined, krone:2013:interactive, guo:2013:coupled, jackson:2013:lightweight, cornel:2015:visualization, schroeder:2014:trend, Lawonn:2016:OBF}      \\    
\bottomrule
\label{table:approach}
\end{tabu}
%\vspace{-2ex}
		\vspace{-1em}
\end{table*}

Finally, \animated\ links can be used. When there is a \transformation\ animation between two representations, one representation transitions to another~\cite{lee:2022:design}, \eg, transitioning a 3D shape model to a diagrammatic 2D counterpart~\cite{miao:2018:dimsum}. Transformations facilitate locating the same data points in multiple representations. When two representations are linked through \modification, interacting with interface components allows viewers to mentally link them.

\subsection{Inherent Relationships between Categories}
Each design dimension is independent, and the design choices within every dimension are formed together to generate a 2D+3D combination. Yet, we found some inherent connections within dimensions. To find supporting evidence, we also analyzed the flow of previous work in picking and combining different design dimensions. We plotted the proportions of design combinations from our literature collection in \autoref{fig:sankey} and, next, list and discuss the links between categories that we observed in previously published approaches.
\begin{itemize}[nosep,left=0pt .. \parindent]
    \item \textbf{\substitute} is linked to \textbf{\positioncn}: When both types of representations are placed in the same view but are rendered at different times, they are typically connected by sharing the same position. This often occurs with \textbf{\transformation} animations because maintaining the positions of the representations highlights correlations between them (\eg, \cite{halladjian:2020:scale, mohammed:2018:abstractocyte}).
    \item A \textbf{\substitute} layout also implies that two representations are unlikely to be \textbf{\interactive} because it is difficult to interact with the visualization while it changes. Usually, the substituting interaction is controlled by another panel, such as a slider, or sometimes by an \abstraction (\eg, \cite{mohammed:2018:abstractocyte,miao:2018:dimsum}).
    \item An \textbf{\embedded} layout implies that the two representations share their \textbf{\positioncn} because the positions in both representations need to be meaningful. For example, 2D representations related to a specific 3D object can be displayed on it~\cite{rocha:2018:illustrative}, and 3D spatiotemporal data can be visualized and located on a 2D map to match the trajectory~\cite{tominski:2012:stacking}.
    \item There are also inherent relationships between using \textbf{\projection} \jh{or \textbf{\slicing}} and visual connections through \textbf{\shapecn}. When a 2D visualization is a projection \jh{of or a slice from a} 3D representation, the structure of the 3D objects remains the same \cite{smit:2017:pelvis, miao:2018:dimsum}, although they are not necessarily in the same orientation \cite{kohler:2018:visual, kretschmer:2013:interactive}. 
    \item Besides, there is also a connection between the two categories of \textbf{\control} and \textbf{\interactive}. If the link is built for the purposes of controlling the representations, then the 2D and 3D representations must necessarily be interactively connected~\cite{kretschmer:2013:interactive, mohammed:2018:abstractocyte}.
    \item \textbf{\control} does not stand on its own as a motivation. To control another representation, the representation must show additional information, \eg, the status of 3D structures \cite{mohammed:2018:abstractocyte} or an overview \cite{glaber:2014:combined}.
\end{itemize}

\subsection{Website}
\label{sec:web}
We implemented a website that allows users to explore the papers we surveyed. The website allows viewers to interactively and easily locate examples of papers that lie within our design space, based on their requirements. Searches can be refined by filtering papers based on their design dimensions and related subcategories. We describe details of our website \jh{in \autoref{app:website}.}
\redsout{and a case study showcasing noteworthy 2D+3D combinations \autoref{app:website} (due to the space limitations of the submission format).}

\section{Design Guidelines}
\label{sec:design_guidelines}
We now describe a set of design guidelines. \jh{We arrived at them by referring to traditional visualization design~\cite{munzner:2014:visualization} and to our design space. We first reflected on how previous work may have arrived at their design with our design space. Then, by analyzing different 2D and 3D representation combinations, we studied potential reasons why authors may have picked such linking approaches. Based on these results, we summarize the following design guidelines.}

\begin{figure*}
    \centering
    \includegraphics[width=\linewidth]{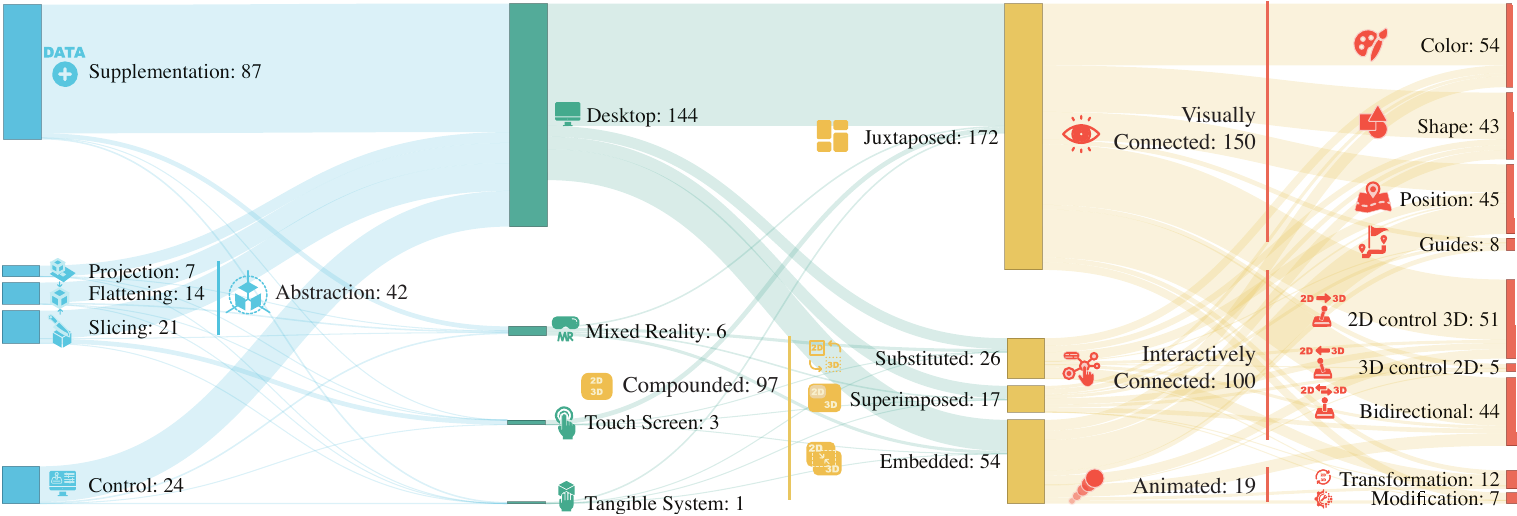}\vspace{-1ex}
    \caption{Alluvial diagram of overall counts in the classes. Note that one paper can be in multiple categories.}
    \label{fig:sankey}
    \vspace{-1.5ex}
\end{figure*}

%\vspace{-0.3ex}
\subsection{Guideline Contents}
\textbf{1. \jh{Define} the motivation behind adding a link between two representations.}
First, the \motivationbg{Motivation} for linking a representation to another should be established. Designers need to consider if an existing representation is able to facilitate the completion of tasks on its own. If it cannot stand on its own, a link between the representation and another that describes supplementary data can be considered (\supplementation). If the visualization suffers from occlusion or is too complex to examine fluently, an abstract representation could be used (\abstraction). If an existing visualization is difficult to navigate or control, a simplified representation can be provided (\control). If possible, designers should leverage existing representations before adding new ones as multiple representations can cause cognitive overhead~\cite{Baldonado:2000:GUM}.

The dimensionality of the existing representation plays an important role in the design of a link. A 2D representation is supplemented by a 3D representation when a designer needs to refer to the spatial structure or add additional data, such as temporal data. A 3D representation can be supplemented by representations that display data that is not naturally encoded in 3D. 2D representations can provide abstractions for 3D representations if the 3D visualization is difficult to examine and if the abstraction has no 3D spatial component. Also, 3D representations can serve as abstractions for other representations when designers want to show simplified 3D structures. \jh{Using a \projection\ can aid viewers in understanding the inner structure of the 3D data, while \flatten\ representations can provide a comprehensive overview. Also, Using a \slicing\ view can aid viewers in understanding the inner structure of the 3D data.}

% consequences of adding unnecessary links
% \marginpar{\small\ti{I removed the development time issue, this is not relevant for a visualization paper I feel, and it sounded like bla, bla}} 
Determining the motivation for adding a link is important, as adding unnecessary links between representations can be detrimental to the system's design~\cite{Baldonado:2000:GUM}. Links while often beneficial, can also increase a system's visual complexity and impact its performance. 

\textbf{2. Determine if the proposed link is best for the current display environment.}
The display environment influences the way the link will be rendered as well as how it is interactively used. A designer should consider how the two representations can be best linked, given the current display environment. In some cases, it may be fruitful to combine multiple display environments together or change the environment altogether. A drastic change in the design, however, should be justified considering the amount of time and resources it may cost to design representations for alternative display environments.

\textit{\displaybg{Digital screens}} (\desktop\ and \touch) are familiar to audiences, easy to develop visualizations for, and simple to interact with. The vast majority of systems we surveyed (97 papers) used this display type for these reasons, but there are situations in which using another type of environment is warranted when facing issues with screen space or when extensible complicated structures are needed. \mr\ can solve these problems, but can be challenging to learn. 
A \tangible\ is straightforward, but changing its data encoding may be difficult. Alternatively, combining multiple environments can be advantageous for certain complex visualization tasks: 

\begin{itemize}[nosep,left=0pt .. \parindent]
    \item \displaybg{Digital Screens} \& \mr: Large 3D spatial datasets that cannot fit on traditional digital screens due to their limited space could benefit from involving \mr~\cite{wang:2020:towards}. An additional advantage of \mr is its ability to naturally encode depth and space, providing an enhanced spatial understanding of the dataset as opposed to \textit{\displaybg{Digital Screens}}.
    \item \displaybg{Digital Screens} \& \tangible: When the 2D+3D combinations are designed for edutainment purposes, a designer may consider using a \tangible\ because playing with physical objects assists users in better understanding~\cite{satriadi:2022:TGD}.
    Such enhanced understanding can apply to both 2D~\cite{suzuki:2019:shapebots} and 3D~\cite{schindler:2020:anatomical} views.
    \item \tangible\ \& \mr: When realistic visual content portrayal and a sense of touch to enhance the data understanding are both important, combinations of both environments could work.
\end{itemize}

\textbf{3. Design an appropriate Layout.} 
The relative positioning, \ie, the \textit{\layoutbg{Layout}} of the representations, is important, as it can build cognitive connections between related objects through their proximity, \ie, \positioncn. \jh{The layout should always emphasize the key elements of each representation and allocate an appropriate amount of screen space according to the importance of each representation. Therefore,} when designing a layout, the designer should consider the data that the two representations share as well as how the layout can facilitate the completion of tasks. If the representations share a small proportion of their data or if they are simultaneously needed for decision-making, they could be placed in \juxtaposed\ views. In this way, the two representations will not occlude each other when completing tasks. When a large amount of data is shared between two representations, \jh{\compounded\ layouts, especially \embedded\ ones, can highlight their common parts. For example, if} the surface of 3D objects needs extra information to be shown,\redsout{we can consider} embedding 2D representations on 3D ones \jh{can explain the detailed information clearly near the target objects} (\eg, \cite{rocha:2018:illustrative}), while when the 2D diagram needs to present other data, such as temporal data,\redsout{we can} \jh{adding 3D representations on it can balance the contexts and supplementary information} (\eg, \cite{tominski:2012:stacking}). 
\jh{Other than these situations, if} it is necessary to keep track of changes between two representations, people can benefit from the \substitute\ layout. \jh{This type of layout may not be appropriate, however, when both representations are necessary for decision-making because switching back and forth between two layout modes creates a high cognitive load.}

\jh{\superimposed\ layouts are often rendered on a much smaller scale in a corner of the display that contains the full-size primary representation. In cases where} one representation provides a non-interactive overview or additional data, such as a mini-map showing the viewer's relative position in a 3D world, \jh{this layout can be effective.}

\redsout{Choosing the appropriate layout is important, as it can significantly affect users in solving their tasks. A \substitute\ layout, for instance, may not be appropriate when both representations are necessary for decision-making; switching back and forth between two layout modes creates a high cognitive load. The layout should always emphasize the key elements of each representation and allocate an appropriate amount of screen space according to the importance of each representation.}

\textbf{4. \redsout{Consider}Connect the two representations visually or through interaction and animations.} 
Layouts may inherently connect two representations through their \positioncn.
To further enhance the link, designers can consider whether the relationship is best expressed visually or via interaction. Visually encoding linked objects is best when there is a limited amount of related objects in both representations: it reduces the cognitive load experienced by the viewer when using the system. Otherwise, interaction that navigates through and filters objects of interest is necessary for viewers to be able to complete their tasks. 

If a designer decides to visually connect two representations, they need to consider the encoding to use. Encoding through \colorcn\ is a simple and powerful way to highlight related objects. However, it can make systems inaccessible to those with difficulties distinguishing colors. Moreover, colors can only encode a limited number of classes before it becomes difficult to distinguish them. Encoding through \shapecn\ is most useful when there is a need to highlight the shape of a 3D structure; in this case, 2D representations can display an outline of the 3D shape. Care should be taken, however, when linking representations through \shapecn\ as viewers often have trouble comparing the area of two shapes. % pretty sure this is in wilkinson's grammar of graphics 
Rendering lines and other marks as \guidecn\ is an intuitive way to connect two representations together. Guides are also applicable in situations when multiple representations supplement one main representation (\eg, \cite{eulzer:2021:visualization}).
Guides, however, can also easily clutter the interface when many related objects are present at once. Moreover, guides often carry the constraint of requiring designers to place two representations next to each other to avoid losing screen space to lines drawn across the screen. Generally, \visual\ views can be powerful, but they often need to be supplemented with interaction to remain effective for a large number of objects.

Interactively connecting two representations makes it easy for viewers to build cognitive links between two representations. Controlling 3D representations with 2D diagrams is most effective in situations where portions of the 3D structure are useful. In this case, the 2D diagram is used to navigate to important portions of the 3D structure. The 2D representation is typically visually connected to the 3D representation to facilitate navigation. A 3D representation can control a 2D diagram that is supplemented with data that does not naturally map to 3 dimensions, such as detailed analytical data about a certain 3D point, \eg, \cite{ye:2021:shuttlespace}. The controlling diagram should be chosen based on the ease of interacting with the primary representation. Bi-directional interactions are best in situations when partial 3D representations with their additional information are needed. In this case, interacting with 3D to update 2D can show corresponding information, while manipulating 2D to control 3D allows users to target the intended 3D data.
A number of factors should be considered when using animations to encode links. A \transformation\ animation requires that the two representations use the same dataset. It works best for cases when the process of reducing or projection a representation is important; \ie, when a viewer needs to keep track of how the process occurred. Modification animations are best when rendering temporal data. Both types of animations can be controlled with interface elements outside of the two representations. 

\vspace{-0.3ex}
\subsection{Case study}
\label{sec:case_study}
\begin{figure*}[t]
    \includegraphics[width=\linewidth]{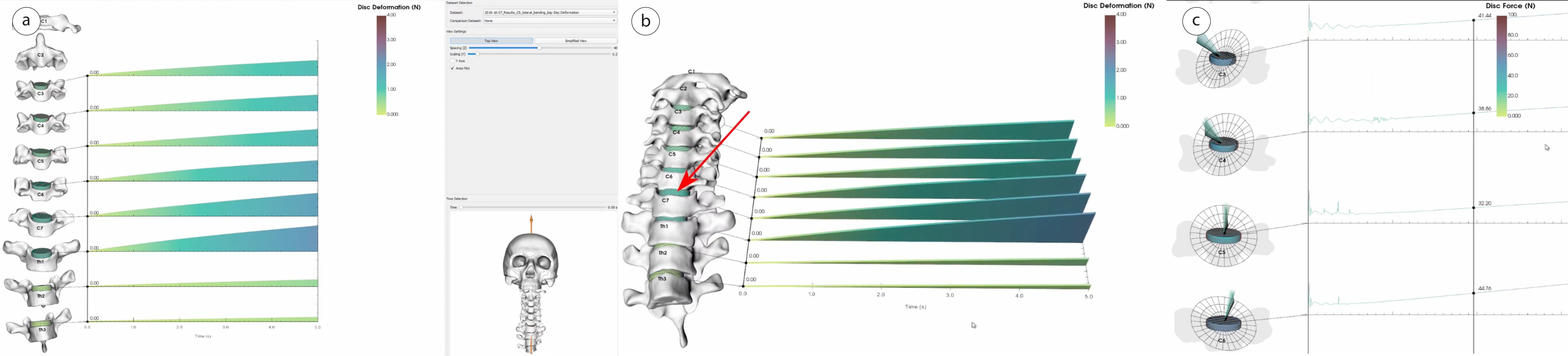}\vspace{-1ex}
    \caption{Screenshots from the system by Eulzer et al.~\cite{eulzer:2021:visualization}: (a) includes the disc deformation in a 2D diagram and corresponding 3D spinal structures; (b) presents the disc deformation with a 3D representation; (c) shows the detailed direction of forces impacting the spinal dices.}
    \label{fig:spnial}
    \vspace{-1.5ex}
\end{figure*}

\jh{To exemplify the practical application of our design guidelines, we analysed a paper authored by Eulzer et al.~\cite{eulzer:2021:visualization} from the perspective of our guideline framework. This paper was chosen because it incorporates multiple 2D and 3D visualizations with various combinations. 
The \motivationbg{Motivation} of integrating 2D and 3D representations lies in the inherent advantages of each: human spine data is effectively showcased in 3D representations, while abstract data, such as disc deformation, is better illustrated in 2D visualizations. Additionally, displaying the moving spinal structures makes it challenging to examine force impact directions. Slicing the structures demonstrates clear relative impact angles. These rationales align with our identified categories---\supplementation\ and \slicing. The choice of implementing the system on a \desktop\ might be due to experts' familiarity with this environment.
The layouts employed for 2D+3D visualizations include \juxtaposed\ (\autoref{fig:spnial}(a)), \superimposed\ (\autoref{fig:spnial}(b)), and \embedded\ (\autoref{fig:spnial}(c)). The 3D animation window in the bottom right corner of \autoref{fig:spnial}(a) requires patient anatomy details, thus favoring the \juxtaposed\ layout for such presentations. Accurately representing the actual force impact necessitates having the direction on the corresponding spinal structures, where \compounded\ layouts are effective.
Regarding the \approachbg{Approaches}, \embedded\ layout inherently combines 3D force impact with 2D anatomy slices through \positioncn. Given that the system includes temporal data to depict dynamic force impact, controlling time in 2D diagrams to update the 3D animation falls into the approach \twocthree. Additionally, as shown in \autoref{fig:spnial}, designers linked 3D disc deformation data with the corresponding anatomy through line \guidecn. Overall, the system enables experts to correlate data with specific anatomical parts effectively.}

\vspace{-0.3ex}
\section{Discussion of Future Directions and Limitations}
In this section, we reflect on our design space and examine future work in developing a design framework for linking 2D and 3D representations, as well as consider examples we found in the literature that were beyond the scope of our paper.

\textbf{Reflecting on the use of 2D and 3D representations.} 
The objective of a visualization is either to aid in the analysis of an object or to present facts and results about an object or experiment~\cite{schulz:2013:DSV}. When building a visualization for analytical purposes, it is essential to include pertinent and comprehensive information. As such, combining 2D and 3D representations may not always be necessary. For instance, experts in some fields may be more used to refer to 2D diagrams, \eg, slices from the microscope, and involving 3D representations can cause extra mental efforts for them~\cite{hong:2022:lineaged}. In other cases where experts primarily explore 3D data, 2D representations can be substituted with data itself~\cite{Mohenska:2022:3ST}. \jh{Overall, designers need to balance the target users' preferences and \motivationbg{Motivations}. The cumulative effects can exceed the summation of the individual type of visual representations in the analysis when (1) the supplemented information is necessary and better shown in the other type of representation; (2) precise abstracted views are crucial; or (3) one representation has multiple statuses to switch, and the other can display the overview and control the statuses.}
When building a visualization that focuses on presenting data, it is important to make it attractive and intuitive. Designers can thus consider building intuitive links based on our design space, \eg, adding modalities other than the visual channel to the representations. For instance, we can use immersive VR~\cite{Hepperle:2019:23S} or tangible interfaces~\cite{suzuki:2019:shapebots} to represent data. 

\textbf{Use of links between 2D and 3D representations.}
We acknowledge that design challenges are inherently complex and can vary significantly across different scenarios. Identifying and validating the optimal link design for a specific case can be a formidable task, which underscores the considerable potential of leveraging various subcategories within a given dimension to address alternative cases. As we detailed in \Cref{sec:web} and in \Cref{app:website}, certain combinations exhibit potential for inspiring future design endeavors. \jh{For instance, adding a cross-section view side-by-side to a complicated 3D structure can help people to select 3D parts occluded from the view~\cite{hadwiger:2012:interactive}. As an alternative, placing the corresponding image view as the background of the 3D structure can provide detailed contexts~\cite{kohler:2018:visual, tominski:2012:stacking}.}
With all these existing examples and enriching literature collections with tags from the design space has the potential to create a robust training dataset for a reinforcement learning model. This model can then provide recommendations on linking 2D and 3D representations in specific cases. 
As discussed previously, unnecessary links between representations can burden viewers and negatively impact task performance. As such, further studies are required to investigate questions such as to what degree 2D+3D combinations should be linked, without being overloaded.

\textbf{Toward a Programmatic Framework for Linking Representations.}
Although there are many techniques designed for creating 2D diagrams (\eg, \cite{bostock:2011:d3}) and 3D models (\eg, \cite{schroeder:1996:toolkit}) separately, there is a lack of tools designed for linking them. Currently, creating a system that integrates linked 2D and 3D representations is labor-intensive, as designers must develop effective designs and manually establish connections. This repetitive process is undertaken by numerous researchers for their specific cases. Representations that can be linked must share part of their data; this fact can be leveraged to develop tools for linking representations. 
For instance, a meta-API for programmers to easily link representations could be developed. Alternatively, a reinforcement learning model can be trained and used, as we mentioned in the previous paragraph, for designers to generate design ideas for proper combinations. 
In this way, we can encourage and inspire more approaches to connect 2D and 3D representations. 

\textbf{Limitations.}
We only included a selection of visualization venues, in particular without those papers published in the human-computer interaction field (\eg, CHI) or specific scientific fields (\eg, cell biology). It is possible that papers in these venues may help extend the design space and reveal new insights. In the future, we plan to investigate a broader range of venues as well. 
Second, our primary focus was on existing literature; however, it is plausible that designers may have identified additional combinations for connecting 2D and 3D representations. Organizing a design workshop could serve to reinforce and expand upon the connections we have established in our examination of these papers. 
Third, while we established the design space through collaborative efforts among all authors \jh{based on our previous work~\cite{Santos:2022:DSL}}, we recognize that further scrutiny and refinement can be achieved by engaging a broader group of experts. \jh{Splitting and summarizing the literature by different facets may benefit in a in-depth understanding of 2D+3D representations.} Also to this end, organizing a workshop with individuals proficient in 2D+3D visual representation design could prove invaluable, as it would allow us to collect feedback to enhance the design space.

\vspace{-1ex}
\section{Conclusion}
In this paper, we explored the design of combinations of 2D and 3D representations based on a survey of published work in the visualization field. 
Echoing the sentiment expressed by Brath~\cite{brath:2015:3d}, we emphasize that the combination of 2D and 3D representations is \emph{here to stay}. Yet rather than merely \emph{dealing with} it, we should actively seek to \emph{take advantage} of it and \emph{embrace} it. By harnessing the design choices offered by our design dimensions, we showcase a multitude of possibilities for seamlessly linking 2D and 3D representations. These representations can be applied not only to 2D/3D data but also to scenarios where data traditionally adheres to \emph{only} 2D or \emph{only} 3D visualization norms. Recognizing the diversity of design opportunities that we found is crucial, and our survey---in conjunction with the accompanying website---can serve as a valuable resource in this regard.

%\vspace{-0.4ex}
\acknowledgments{We thank the authors we contacted for their support in clarifying their papers and the Vader team for their general comments. The work was partially supported by the U.S. Department of Homeland Security under Grant Award Number 17STQAC00001-07-00.}

%\vspace{-0.4ex}
\section*{Images/graphs/plots/tables copyright}
The images shown in Figures~\ref{fig:motivations}--\ref{fig:approach} and \ref{fig:spnial} are all from papers published by the IEEE Computer Society and most are copyrighted by the IEEE, we use them here with permission. \jh{The remaining images in these figures are used under the \href{https://creativecommons.org/licenses/by/4.0/}{Creative Commons At\-tri\-bu\-tion 4.0 International (\ccLogo\,\ccAttribution\ \mbox{CC BY 4.0})} license: \autoref{fig:layout}(b), \ref{fig:layout}(c), and \ref{fig:approach}(c).}
The remaining Figures~\ref{fig:overview_design_space} and \ref{fig:sankey} and all tables are and remain \textcopyright\ by this paper's authors and are used under the \href{https://creativecommons.org/licenses/by/4.0/}{Creative Commons At\-tri\-bu\-tion 4.0 International (\ccLogo\,\ccAttribution\ \mbox{CC BY 4.0})} license (shared at \href{https://osf.io/rhygf/}{\texttt{osf.io/rhygf}}).

\bibliographystyle{abbrv-doi-hyperref}
\bibliography{template}

\begin{figure*}[b!]
	%\vspace{-.25ex}%
	\centering%
	\includegraphics[width=\linewidth]{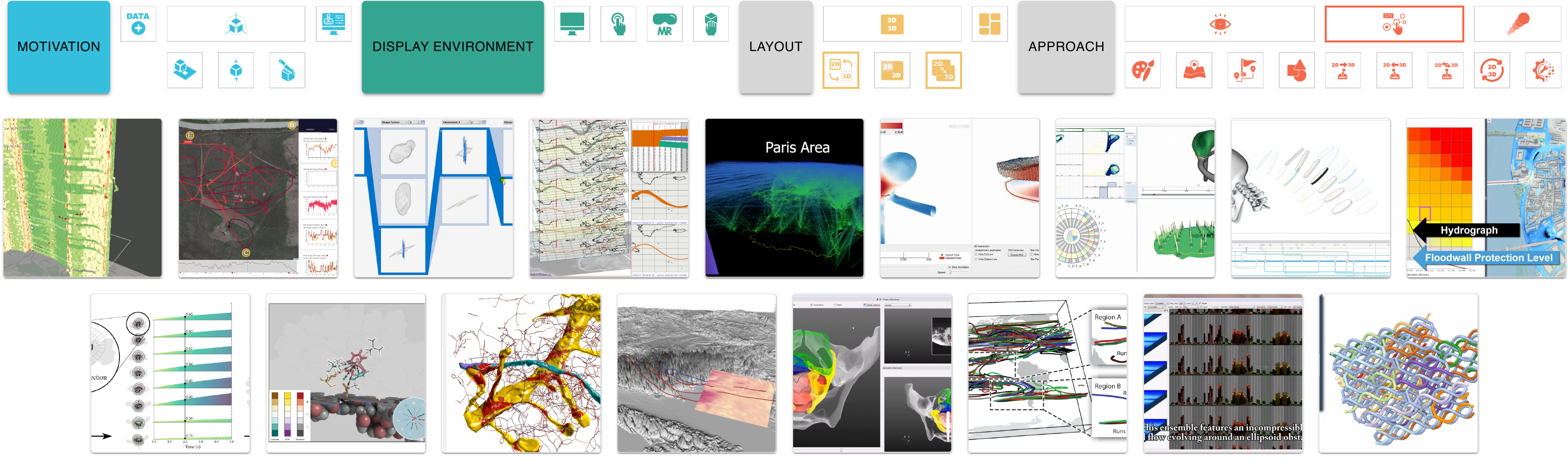}\vspace{-1ex}
	\caption{An example of using the website: filtering for \substitute\ and \embedded\ layout and \interactive.}
	\label{fig:website_filter}
\end{figure*}

\begin{figure*}[b!]
    \centering
    \includegraphics[width=\linewidth]{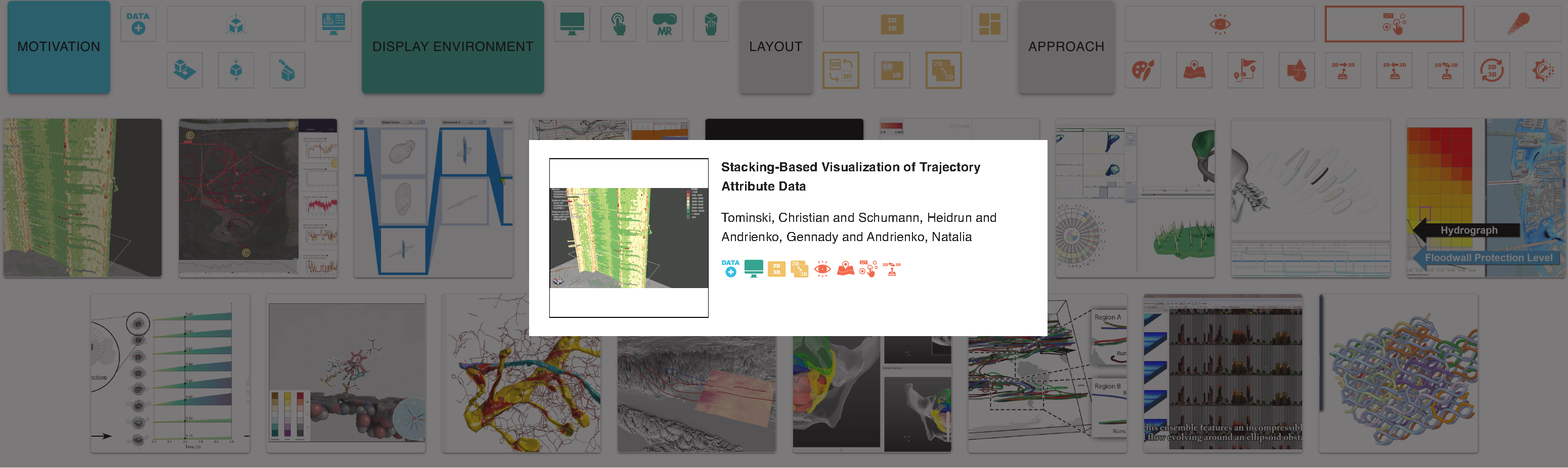}\vspace{-1ex}
    \caption{Screenshot with details about the paper by Tominski et al.~\cite{tominski:2012:stacking}.}
    \label{fig:detailed_paper_view}
\end{figure*}

\newpage

\begin{strip} % requires \usepackage{cuted}
\noindent\begin{minipage}{\textwidth}
\makeatletter
\centering%
\sffamily\bfseries\fontsize{15}{16.5}\selectfont
\vgtc@title\\[.5em]
\large Appendix\\[.75em]
\makeatother
%\normalfont\rmfamily\normalsize\noindent\raggedright In this appendix we provide material beyond the what we could include in the main paper due to space limitations. In particular, note that we have more than 3.5 pages of references to be able to cite all papers that we survey, so have less space overall in the main paper to describe things such as the website and some noteworthy 2D+3D combinations in detail, so we add this detail here in the appendix.
\end{minipage}
\end{strip}

\appendix

\begin{figure*}[t!]
    \centering
    \includegraphics[width=0.94\linewidth]{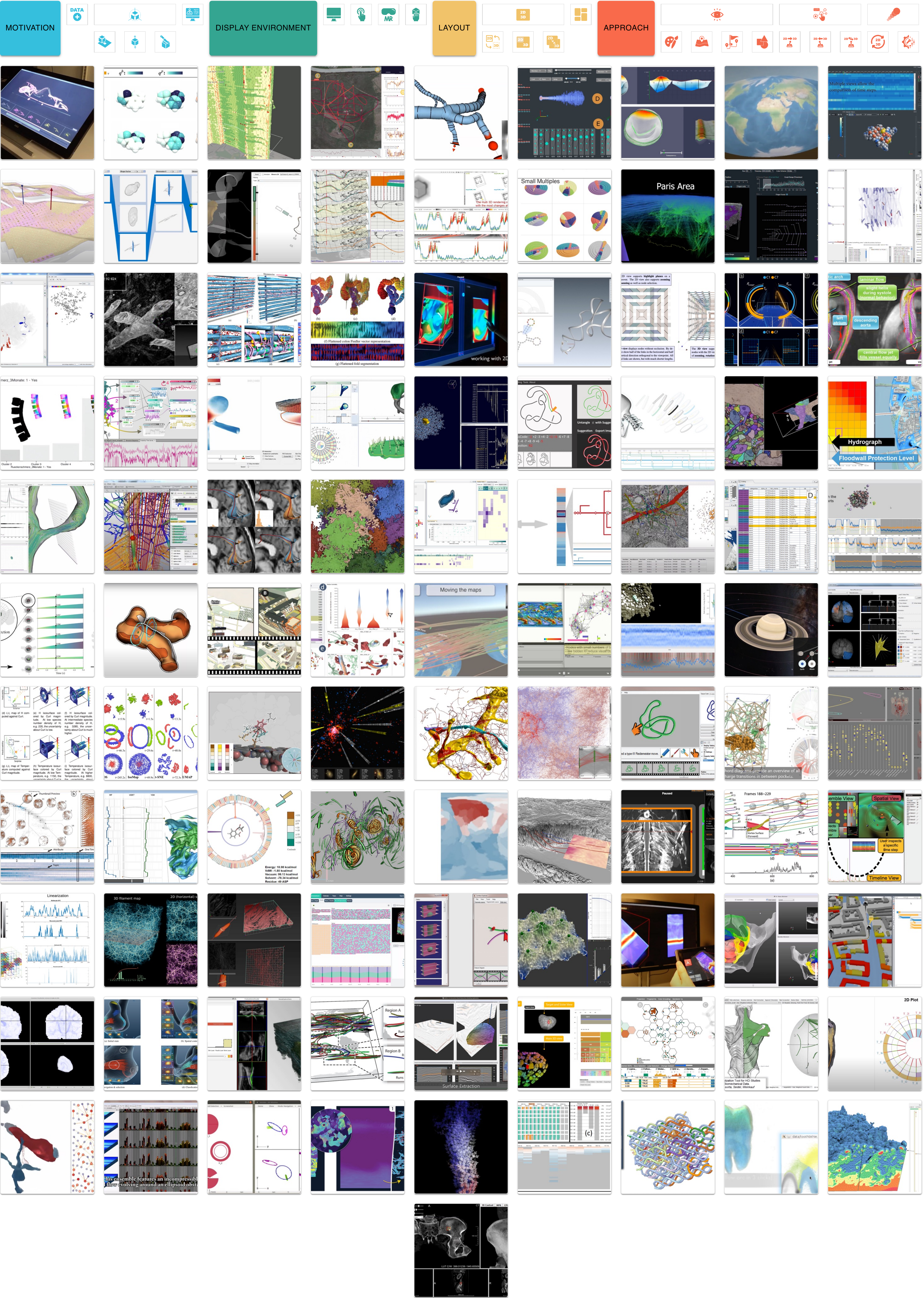}\vspace{-1ex}
    \caption{Screenshot our website without any filtering, showing an overview of all papers.}
    \label{fig:website}
\end{figure*}

\section{Website}
\label{app:website}
\subsection{Website Interaction}
Our website \jh{(\href{https://2dplus3d.github.io/}{\texttt{2dplus3d\discretionary{}{.}{.}github\discretionary{}{.}{.}io}})}
% \marginpar{\todo{placeholder: will update the link later---if this is actually a short github page that would be best}}
shows our classification of all \jh{105} papers we identified in our survey that contain examples for how to link 2D with 3D data views. We initially present an icon-based overview of the papers, as we illustrate in \autoref{fig:website}. From there, people may wish to focus on a specific subcategory, \eg, \embedded\ layout. This selection can be easily accomplished by means of the corresponding icon at the top of the page to engage a corresponding filter. If people are interested in multiple layout types, they have the flexibility to select multiple options for an OR-based filter combination. Furthermore, to conduct a cross-dimensional analysis, they can pick additional features in which they are interested from another category (for an AND-based combination), \eg, \interactive. This selection then results in a view as we show in \autoref{fig:website_filter}. To access detailed information about a given specific paper, people can simply click on its icon (see an example in \autoref{fig:detailed_paper_view}). Then we show details about the publication and provide a direct link to the digital library entry of the original paper via the paper's official DOI.

\subsection{Implementation}
The website for our survey was generated using a Python framework we built specifically for generating websites for survey papers (\href{https://github.com/VADERASU/survey_framework}{\texttt{github\discretionary{}{.}{.}com\discretionary{/}{}{/}VADERASU\discretionary{/}{}{/}survey\_\discretionary{-}{}{}framework}}). Authors can generate their survey paper websites with the framework we built. 
They simply need to supply a metadata file, a LaTeX bibliography file (\texttt{*.bib}), as well as directories containing icons and image data. An extraction script utilizes all these aforementioned data to generate a front-end and database.
It generates a MongoDB database, which is accessible through a front-end powered by React by running the FastAPI back-end server we included. 
As such, this framework can facilitate the addition of future 2D+3D combination designs for our website over the long term.

\section*{Images copyright}
Figures~\ref{fig:website_filter}--\ref{fig:website} \jh{are screenshots of our website in use} and remain \textcopyright\ by this paper's authors and are used under the \href{https://creativecommons.org/licenses/by/4.0/}{Creative Commons At\-tri\-bu\-tion 4.0 International (\ccLogo\,\ccAttribution\ \mbox{CC BY 4.0})} license (shared at \href{https://osf.io/rhygf/}{\texttt{osf.io/rhygf}}).

\end{document}